\documentclass[12pt]{article}
\usepackage{epsf}
\makeatletter \@addtoreset{equation}{section} \makeatother

\def\cA{{\cal A}} \def\cB{{\cal B}}
 \def\cD{{\cal D}}
 
\def\cF{{\cal F}} 
 
 \def\cK{{\cal K}}
\def\cL{{\cal L}} \def\cM{{\cal M}}
\def\cN{{\cal N}} 
 
\def\cR{{\cal R}} \def\cV{{\cal V}}
 
\def\cU{{\cal U}}
\def\trace{\mathop{\rm Tr}\nolimits}
\newcommand{\dfrac}{\displaystyle \frac}
\def\rmi{{\rm i}}
\def\rmd{{\rm d}}
\newcommand{\ft}[2]{{\textstyle\frac{#1}{#2}}}
\def\tx{x}
\def\ty{y}
\newsavebox{\uuunit}
\sbox{\uuunit}
    {\setlength{\unitlength}{0.825em}
     \begin{picture}(0.6,0.7)
        \thinlines
        \put(0,0){\line(1,0){0.5}}
        \put(0.15,0){\line(0,1){0.7}}
        \put(0.35,0){\line(0,1){0.8}}
       \multiput(0.3,0.8)(-0.04,-0.02){12}{\rule{0.5pt}{0.5pt}}
     \end {picture}}
\newcommand {\unity}{\mathord{\!\usebox{\uuunit}}}
\def\Re{\mathop{\rm Re}\nolimits}

\def\diag{\mathop{\rm diagonal}\nolimits\ }
\def\eig{\mathop{\rm eigenvalues}\nolimits\ }
\begin{document}

\begin{titlepage}

\begin{flushright}
HU-EP 01/12\\
SU-ITP-01/13\\
KUL-TF-01/07\\
CITUSC/01-012\\
hep-th/0104056
\end{flushright}

\begin{center}

{\LARGE \bf Hypermultiplets, domain walls and

\medskip

supersymmetric attractors}

\vspace{0.5 cm}

{\Large Anna Ceresole$^{\star}$, Gianguido Dall'Agata$^{\sharp}$,\\[3mm]
Renata Kallosh$^\natural$ and Antoine Van Proeyen$^{\flat}$}

\vspace{0.5 cm}

{ $\star$ \it Dipartimento di Fisica, Politecnico di Torino and \\
Istituto Nazionale di Fisica Nucleare, Sezione di Torino \\
Corso Duca degli Abruzzi, 24, I-10129 Torino\\
{\tt ceresole@athena.polito.it}}

\medskip

{$\sharp$ \it Institut f{\"u}r Physik, Humboldt Universit{\"a}t \\
Invalidenstra\ss{}e 110, 10115 Berlin, Germany\\
{\tt dallagat@physik.hu-berlin.de}}

\medskip

{$\natural$ \it Department of Physics, Stanford University, \\
Stanford, CA 94305, USA\\
{\tt kallosh@stanford.edu}}

\medskip

{$\flat$ \it Instituut voor Theoretische Fysica, Katholieke Universiteit Leuven \\
Celestijnenlaan 200D, B-30001 Leuven, Belgium\\
{\tt Antoine.VanProeyen@fys.kuleuven.ac.be}}
\end{center}
\vspace{0.5 cm}

\begin{abstract}

We establish general properties of supersymmetric flow equations and of
the superpotential of five-dimensional $\cN = 2$ gauged supergravity
coupled to vector multiplets and hypermultiplets. We provide necessary and
sufficient conditions for BPS domain walls and  find a set of algebraic
attractor equations for $\cN = 2$ critical points.

As an example we describe in detail the gauging of the universal
hypermultiplet and a vector multiplet. We study a two-parameter family of
superpotentials with supersymmetric AdS critical points and we find, in
particular, an $\cN = 2$ embedding for the UV--IR solution of Freedman,
Gubser, Pilch and Warner of the $\cN = 8$ theory. We comment on the
relevance of these results for brane world constructions.
\end{abstract}

\vskip 10mm

Published in Phys. Rev. {\bf D64} (2001) 104006
\end{titlepage}

\newpage

\baselineskip 6 mm

\section{Introduction and goals}
Among the remarkable spinoffs of the AdS/CFT correspondence between
strings on $AdS\times X$ and boundary superconformal theories, a lot of
interest is devoted to the duality between domain wall supergravity
solutions and renormalisation group (RG) flows of field theory couplings.

The purpose of this paper is to find out the general properties of
supersymmetric flows and vacua of $D=5$, $\cN=2$ supergravity coupled to
hypermultiplets and vector multiplets with nonconstant scalars, based on
the theory of~\cite{AnnaGianguido}. This analysis aims at following the
lines of the attractor flows~\cite{attractflows} for black holes in four
and five dimensions that paved the way to finding
Bogomol'nyi--Prasad--Sommerfield (BPS) solutions.

The basic interest in domain wall or supersymmetric flows is due to the
possible correspondence between BPS domain walls of gauged supergravity
and exact supersymmetric vacua of the fundamental M or string theory.
Unlike black holes, the domain wall  solutions in $D=5$ interpolating
between AdS vacua do not break $D=4$  Lorentz symmetry and therefore may
give interesting possibilities for the realistic vacua of our 4D world.
On the other hand, the intuition gained in studies of black hole
attractors may be useful for understanding the issues of stabilization of
moduli at supersymmetric vacua.

The gauging of supergravity in the vector multiplet sector has been
studied with respect to the supersymmetric vacua of the theory. In
particular, for U(1) $D=5$ gauged supergravity the supersymmetric vacua
are defined by the superpotential $W=h^I(\phi)V_I$, which has a
dependence on moduli analogous to the black hole central charge
$Z=h^I(\phi)q_I$. Here $h^I(\phi)$ are special coordinates, $V_I$ are
constants related to the Fayet--Iliopoulos terms, and $q_I$ are black hole
electric charges. Many critical points of both these systems are known.
One always finds AdS${}_5$ vacua (and domain walls with AdS vacua) when
scalars from vector multiplets reach their fixed points. These fixed
points are specified by the algebraic equation $V_I = h_I(\phi_{\rm
cr})W_{\rm cr}$, where $h_I(\phi)$ are the dual special coordinates. This
equation is analogous to the equation $q_I = h_I(\phi_{\rm cr})Z_{\rm
cr}$, which defines the fixed scalars near the horizon of the $D=5$
electrically charged black holes. Both of these equations were derived
and analysed in~\cite{Chou:1997ba}.

The solutions without hypermultiplets are also known to have specific
properties, like the fact that they always approach a UV fixed point,
i.e., the AdS boundary~\cite{KL,BC}. This feature leads to a no--go
theorem for the `alternative to compactification' Randall--Sundrum (RS)
smooth scenario~\cite{RS}.

However, the situation may change when hypermatter is added and thus it
is important to elucidate the nature of the fixed scalars in the BPS
domain wall configurations in this case. The first examples of domain
walls were found in the coupling with the universal hypermultiplet. The
one in~\cite{Lukas:1998tt} does not have AdS critical points, whereas the
one in~\cite{DW} displayed one UV and one IR critical point. In the
latter, the authors claimed that their model could give an $\cN =2$
realization of the domain wall solution found by Freedman, Gubser, Pilch
and Warner~\cite{FGPW} (FGPW) as holographic dual to a RG flow from an
$\cN = 4$ to an $\cN = 1$ Yang--Mills theory. However, this description
relied on a nonstandard formulation of 5D supergravity which has not been
proven to be consistent.

Other U(1) gaugings of the same model were recently studied
in~\cite{Falkowski:2000yq,Behrndt:2001qa,Behrndt:2000ph}.

This paper starts with a systematic description of supersymmetric flow
equations in the presence of vector and hypermultiplets. We first solve
the issue of describing both the flows and the attractor points in terms
of a single superpotential $W$ for generic (non--Abelian) gaugings. This
is nontrivial, since the theory is defined in terms of the SU(2) triplet
of quaternionic prepotentials $P_I^r(q)$ dressed with the $h^I(\phi)$
functions of the scalars in vector multiplets.

We will find that {\it the single superpotential $W(\phi, q)$ is related
to the norm of the dressed prepotentials $P^r(\phi ,q)$ and controls the
supersymmetric flow equations if and only if their SU(2) phase satisfies
the constraint}
\begin{equation}
\partial_{\phi} Q^r =0\, \quad\mbox{where}\quad P^r(\phi , q)\equiv h^I(\phi ) P_I^r(q)\equiv  \sqrt {2/3} \, W Q^r\,,
\qquad (Q^r)^2=1 \,. \label{phase}
\end{equation}
This may restrict the class of gauged supergravities with hypermultiplets
that have BPS solutions.

We then characterize the critical points by a set of attractor equations.
It parallels the attractor mechanism for moduli near the black hole
horizon~\cite{attractflows} and is also supported by an enhancement of
unbroken supersymmetry near the AdS vacua. The attractor equations are
simple algebraic conditions which fix the values of the moduli:
\begin{equation}
P^r_I(q_{\rm cr}) - h_I(\phi_{\rm cr}) P^r_{\rm cr}(\phi_{\rm cr},q_{\rm
cr})=0\ , \qquad K^X_{\rm cr} \equiv h^I(\phi_{\rm cr}) K_I^X(q_{\rm
cr})=0 \,. \label{attr}
\end{equation}
The first equation is defined by the very special geometry and is
analogous to the one for black holes discussed above. The other one
requires a certain combination of quaternionic Killing vectors to
vanish.\footnote{This requirement appeared in~\cite{AnnaGianguido} and was
then noticed in~\cite{Behrndt:2001qa} for BPS instantons and also
in~\cite{D'Auria:2001kv}.} These algebraic equations are very useful in
simplifying the general analysis of critical points, as they replace the
differential equations for the extrema of the superpotential. They will
prove to be very useful also in our simple cases.

Our  general theory will be applied to the model of the universal
hypermultiplet alone as well as coupled to one vector multiplet. The full
moduli space is
\begin{equation}
   \cM = \mathrm{O(1,1)} \times \frac{\mathrm{SU(2,1)}}{\textrm{SU(2)}\times \textrm{U(1)}}\,.
\end{equation}

For the hypermultiplet alone, we study the properties and
parametrizations of the scalar manifold
$\frac{\mathrm{SU(2,1)}}{\mathrm{SU(2)}\times \mathrm{U(1)}}$, giving all
Killing vectors and prepotentials that allow us to write down the generic
scalar potential. Analyzing all U(1) gaugings systematically, we find
that only one critical point arises. On general grounds we give precise
conditions for determining its (UV/IR) nature, which interestingly can be
tuned by the choice of the direction gauged within the compact subgroup.
We compare results with other parametrizations, that, due to an
ill-defined metric, can give rise to spurious singular points.
Specifically, in appendix~\ref{app:BC}, we show how this happens
in~\cite{Behrndt:2000ph}, where the parametrization
of~\cite{Britto-Pacumio:1999sn} was used.

Then we turn to the full model, where we analyse the most general
$\textrm{U(1)}\times \textrm{U(1)}$ gauging. The requirements for a first
critical point lead to three real parameters for the embedding in
$\textrm{SU(2)}\times \textrm{U(1)}$. A linear relation determines
whether extra possibilities exist where noncompact generators of SU(2,1)
contribute to one of the U(1) generators. We further study a 2-parameter
subclass. We then restrict ourselves to the theories that have 2
different AdS critical points, as they could be extrema of RG-flows as
well as of Randall--Sundrum type smooth solutions. This leaves us with
only 2 independent real numerical parameters $\beta $ and $\gamma $. The
superpotential of the 2-parameter model existing on a line in the
quaternionic manifold
parametrized by $\chi$ is 
\begin{equation}
  W =  \frac{1}{4\,\rho^{2}} \left[3 + \,\beta +\left( \ft32 + \gamma\right)\,\rho^6 +
    \left( 1 - \beta + (\ft12 - \gamma)\,\rho^6 \right)
       \,\cosh (2\,\chi)\right] ,
\end{equation}
where $\rho$ is the vector modulus.

Again, the nature of the critical points depends on the relation between
the parameters. Quite remarkably, for the special values $\beta =-1$ and
$\gamma =3/2$ we recover precisely the superpotential and the UV--IR AdS
critical points of the kink solution of Freedman, Gubser, Pilch, and
Warner~\cite{FGPW}. This means that the  FGPW flow can be described all
within $D=5$, $\cN = 2$ gauged  supergravity \cite{AnnaGianguido} coupled
to one vector and one hypermultiplet, and thus the corresponding sector
of the $\cN = 2$ theory yields a consistent truncation of ten-dimensional
type IIB supergravity~\cite{10IIB}.

As an outcome of this analysis, we find as an interesting feature that
domain walls with hypermultiplets can give rise to IR directions. This
removes in principle the main obstacle for realizing a smooth
supersymmetric RS scenario with a single brane (RSII), which in the case
of vector multiplets only, resulted in the no-go theorem of~\cite{KL,BC}.
The next step would be to find in a specific model 2 IR critical points
and an interpolating solution such that the warp factor obtains a
maximum. The supergravity flow equations impose the condition that $A' =
\pm W$ and thus the existence of such a maximum implies a vanishing
superpotential.

On the other hand, the holographic $c$
theorem~\cite{GPPZ2,FGPW,Skenderis:1999mm} imposes the monotonicity of the
$c$ function $c \sim |W|^{-3}$. This should imply that it is impossible to
connect smoothly 2 IR points and find a smooth supersymmetric RSII.
However, the only condition imposed by supergravity and by the BPS flow
equations is the ``monotonicity theorem'' $A''\leq 0$. This leads to a
monotonicity of the first derivative of the warp factor $A' = \pm W$ and
not in general of the $c$ function. This does not exclude flows where the
superpotential reaches $W=0$. These points may signal some problem with
the validity of the five--dimensional supergravity approximation of the
holographic correspondence~\cite{Porrati:2000nb}. However, BPS flows
crossing such points are perfectly well behaved from the supergravity
perspective. In our specific study we find examples with $W$ vanishing at
some points,\footnote{In~\cite{Gibbons:2000hg} it is shown that a
world-volume theory for a domain--wall at such a place has problems due
to unbounded fermions. This has been investigated in the context of
theories with only vector multiplets. It should be investigated whether
similar problems persist for fermions with transformation laws like those
in hypermultiplets.} but the flows by these points always lead to a naked
singularity. We conclude therefore that, although no example exists at
the moment, a smooth supersymmetric realization of RSII does not seem to
be ruled out in the presence of vector and hypermultiplets.

Conversely, it is likely that a realistic one--brane Randall--Sundrum
scenario can be constructed on the basis of any of the models discussed
above with at least one IR critical point by employing a method with
supersymmetric singular brane sources~\cite{susyd5sing}.

The paper proceeds in section~\ref{ss:flowtools} with a general
discussion of supersymmetric flows with an arbitrary number of vector and
hypermultiplets based on the most general consistent gaugings. We start
by repeating the general ingredients of the very special real and the
quaternionic manifolds, and the gauging of the isometries. We provide a
general constraint for gravitational stability that can be expressed as
the BPS condition in a domain wall background. We spell out the
requirements for a RSII scenario in terms of the concepts of a
renormalization group flow. The requirement for critical points of the
(super)potential can be reduced to algebraic conditions, the attractor
equations. We end this section with a summary of the features to be
investigated in examples.

Section~\ref{ss:scManIsom} starts with motivations for studying a simple
model with a vector multiplet and a hypermultiplet such as the one giving
rise to the ${\cal N}=2$ description of the FGPW flow. Then we study
properties and parametrizations of the scalar manifold, giving all
Killing vectors and prepotentials that allow us to write down the generic
scalar potential.

Section~\ref{ss:GaugingFlow} provides an analysis of gauges and flows in
examples. For the toy model with only a universal hypermultiplet, only one
critical point arises, and, depending on the direction gauged within the
$\textrm{SU(2)} \times \textrm{U(1)}$ compact subgroup, it can have
different nature. Then we study the full model and examine the
possibility of finding different flows between two fixed points, proving
in particular that the FGPW flow can be recovered.

We finish with some concluding remarks in section~\ref{ss:conclRem}.

Our conventions are generally those of~\cite{AnnaGianguido}. In
appendix~\ref{app:indices} we present a convenient table for the reader
to recall the use and range of all the indices. In
appendix~\ref{app:reality} we repeat some notational issues, paying
attention to reality conditions. In appendix~\ref{app:BC} we comment on
the toy model in a different parametrization, useful for comparison
with~\cite{Behrndt:2000ph}.

\section{Supersymmetric flow equations and \\
domain wall attractors} \label{ss:flowtools}
\subsection{Basic aspects of the theory}
The bosonic sector of 5D, $\cN=2$ supergravity coupled to $n$ vector
multiplets and $r$ hypermultiplets\footnote{We omit tensor multiplets for
simplicity.} has as independent fields the f{\"u}nfbein $e_\mu ^a$, the $n+1$
vectors $A_\mu ^I$ with field strengths $F_{\mu \nu }^I= \partial _\mu
A_\nu ^I-\partial _\nu A_\mu ^I+gA_\mu ^JA_\nu ^K f_{JK}{}^I$, the $n$
scalars $\phi ^{\tx}$, and the $4r$ ``hyperscalars'' $q^X$. Full results
of the action and transformation laws are in \cite{AnnaGianguido}. We
repeat here the main ingredients (for some technical issues, see
appendix~\ref{app:reality}). The bosonic part of the Lagrangian is
\begin{eqnarray}
e^{-1} \cL^{\cN = 2}_{\rm bosonic} &=& - \ft{1}{2} R
-\ft{1}{4}a_{IJ}F_{\mu \nu }^I F^{J\mu \nu }
-\ft{1}{2} g_{XY} \cD_\mu q^X \cD^\mu q^Y  - \ft{1}{2}
g_{\tx\ty} \cD_\mu \phi^{\tx} \cD^\mu \phi^{\ty} +\nonumber\\
&&+\frac{1}{6\sqrt{6}} C_{IJK}
e^{-1}\varepsilon^{\mu\nu\rho\sigma\tau} F_{\mu\nu}^I
F_{\rho\sigma}^J A^K_\tau
 - g^2 \cV(\phi,q)\,, \nonumber
\end{eqnarray} where
\begin{equation}
{\cal D}_\mu q^X=\partial _\mu q^X+gA_\mu ^IK_I^X(q)\,, \qquad {\cal
D}_\mu \phi ^{\tx}=\partial_\mu \phi ^{\tx}+gA_\mu ^IK_I^{\tx}(\phi )\,.
 \label{calDbos}
\end{equation}
Here $K^X_I(q)$ are the Killing vectors of the gauged isometries on the
quaternionic scalar manifold parametrized by the hyperscalars $q^X$,
whereas $K^{\tx}_I(\phi )$ are those of the very special manifold spanned
by the $\phi^{\tx}$ of the vector multiplets. We will come back to these
below.

The scalars of the vector multiplets can be described by a hypersurface
in an $(n+1)$-dimensional space \cite{Gunaydin:1984bi}
\begin{equation} C_{IJK} h^I(\phi ) h^J(\phi ) h^K(\phi )=1 \label{Chhh1}\,.
\end{equation}
The real coefficients $C_{IJK}$ determine the metrics of ``very special
geometry'' \cite{brokensi}
\begin{eqnarray}
   a_{IJ}&\equiv &-2C_{IJK}h^K+3C_{IKL}C_{JMN}h^Kh^Lh^Mh^N=  h_I h_J + h_{\tx I}h^{\tx}_J\,, \nonumber\\
 g_{\tx\ty} & \equiv  & h_{\tx}^Ih_{\ty}^J a_{IJ}\,, \qquad
h_I= C_{IJK}h^Jh^K\,,\qquad h_{\tx}^I\equiv  -\sqrt{\ft32} \partial
_{\tx}h^{I}(\phi )\,,
 \label{defag}
\end{eqnarray}
which are further used for raising and lowering indices. A
non-Abelian structure in the absence of tensor multiplets should
satisfy
\begin{equation}
 {\phantom{f}C}_{L(IJ}f^L_{K)M}=0\,,\qquad K_I^{\tx} =
\sqrt{\ft32}h_Kf^K_{JI}h^{J{\tx}}
 \label{constraintgauging}
\end{equation}
which implies
\begin{equation}
  h_If^I_{JK}h^K=0\ \rightarrow K_I^{\tx}h^I=0\,.
 \label{hfh0}
\end{equation}

The quaternionic K{\"a}hler geometry is determined by $4r$-beins $f^{iA}_X$
(as one-forms $f^{iA}=f^{iA}_X dq^X$), with the SU(2) index $i=1,2$ and
the Sp(2$r$) index $A=1,\ldots,2r$, raised and lowered by the symplectic
metrics $C_{AB}$ and $\varepsilon_{ij}$ (see appendix~\ref{app:reality}
for conventions, reality conditions, etc). The metric on the hyperscalar
space is given by
\begin{equation}
  g_{XY}\equiv  f_{X}^{iA}f_{Y}^{jB}\varepsilon
  _{ij}C_{AB}=f_{X}^{iA}f_{YiA}\,.
 \label{defgXY}
\end{equation}
This implies that the vielbeins satisfy also
\begin{equation}
  f_{iA}^X f^{iA}_Y=\delta _Y^X\,,\qquad  f^X_{iA}f^{jB}_X=\delta _i{}^j\delta
  _A{}^B\, .
 \label{ff}
\end{equation}
They are  covariantly constant, including Levi--Civit\`{a} connection $\Gamma
_{XZ}{}^Y$ on the manifold, Sp(2$r$) connection $\omega _X{}^B{}_A$, and
SU(2) connection $\omega _{Xi}{}^j$, which are all functions of the
hyperscalars:
\begin{equation}
\partial _X f_Y^{iA}-\Gamma _{XY}{}^Z f_Z^{iA}
+f_Y^{iB}\omega  _{XB}{}^A
  +\omega _{Xk}{}^i f_Y^{kA}=0\,. \label{covconstf}
\end{equation}
The SU(2) curvature is
\begin{equation}
  {\cal R}_{XYij}= f_{XC(i}f_{j)Y}^C \,,
 \label{RSU2f}
\end{equation}
and there is a connection $\omega _{\mu i}{}^j=(\partial _\mu q^X) \omega
_{Xi}{}^j$ such that
\begin{equation}
  {\cal R}_{XYi}{}^j=2\partial _{[X}\omega_{Y]i} {}^{j}-2\omega_{[X|i|}{}^k\omega
  _{Y]k}{}^j=\rmi {\cal R}^r_{XY}(\sigma _r)_i{}^j\,,\qquad
\cR^r = d \omega^r - \varepsilon^{rst} \omega^s \omega^t
 \label{cRXYij}
\end{equation}
with $r=1,2,3$ and real ${\cal R}^r_{XY}$ [see (\ref{RijRr})]. The SU(2)
curvatures have a product relation that reflects that they are
proportional to the three complex structures of the quaternionic space
\begin{equation}
  {\cal R}^r_{XY}{\cal R}^{sYZ}=-\ft14\delta ^{rs}\delta _X{}^Z
   -\ft12\varepsilon ^{rst}{\cal R}^t_X{}^Z\,.
 \label{algRSU2}
\end{equation}

The Killing vectors on the hyperscalars $K_I^X$ can be obtained from an
SU(2) triplet of real prepotentials $P_I^r(q)$ that are defined by the
relation~\cite{Galicki:1987ja,DFF,D4N2,AnnaGianguido}:
\begin{equation}
{\cR}_{XY}^r K_I^Y= D_X P_I^r \,,\qquad D_X P_I^ r\equiv
\partial _XP_I^ r+2\varepsilon ^{rst}\omega _X^sP_I^t \,.\label{defprep}
\end{equation}
These yield [using (\ref{algRSU2})]
\begin{equation}
K^Z_I=-\ft43 {\cal R}^{r\,ZX}  D_X P_I^r\,.
 \label{KXandDP}
\end{equation}
These prepotentials satisfy the constraint
\begin{equation}
   \ft{1}{2}{\cal
R}^r_{XY}K^X_IK^Y_J-\varepsilon^{rst}P^s_IP^t_J+\ft{1}{2}f_{IJ}^K
 P^r_K =0\,.
 \label{constr2prep}
\end{equation}
In local supersymmetry, the prepotentials are defined uniquely from the
Killing vectors. Indeed,\footnote{This formula can also be derived from
the harmonicity property of the quaternionic prepotential $D^X D_X
P^r_I=2r P^r_I$ \cite{D'Auria:2001kv}.}
\begin{equation}
  P^r_I=  \frac{1}{2r}D_XK_{IY}{\cal R}^{XYr}
 \label{valuePrI}
\end{equation}
satisfies (\ref{defprep}), and any covariantly constant shift
$P^{(0)r}_I$ is excluded as the integrability condition $\varepsilon
^{rst} {\cal R}^s_{XY}P^{(0)t}=0$ implies that $P^{(0)r}_I=0$. As in four
dimensions~\cite{D4N2}, these shifts are interpreted as the analogues of
the Fayet--Iliopoulos (FI) terms for the $D = 4$, $\cN = 1$ theories.
However, in local supersymmetry we thus find the absence of the FI term
except when there are no hypermultiplets [or in rigid supersymmetry where
the SU(2) curvature vanishes].

We will also need the bosonic part of the supersymmetry transformations of
the fermions, which are (with vanishing vectors)
\begin{eqnarray}
\delta _\epsilon \psi _{\mu i}  & = & { D}_\mu (\omega )\epsilon _i+\rmi
\frac{1}{\sqrt{6}}g\gamma _\mu P_{ij}\epsilon ^j=
\partial  _\mu  \epsilon _i+ \ft14 \gamma^{ab} \omega_{\mu ,
ab}\epsilon _i  
- \omega_{\mu i} {}^{j}\epsilon_j +\rmi
\frac{1}{\sqrt{6}}g\gamma _\mu P_{ij}\epsilon ^j\,, \nonumber\\
 \delta _\epsilon \lambda _i^{\tx}&=&-\rmi \ft{1}{2}(\not\!{\partial} \varphi
^{\tx})\epsilon _i
+g \epsilon ^jP_{ij}^{\tx}\,,\nonumber\\
 \delta_\epsilon \zeta^A &=& -\rmi\ft{1}{2} f_{iX}^A(\not\!{\partial}
  q^{X})\epsilon ^i + g\epsilon ^i{\cal N}_i^A\,,
 \label{transfos1}
\end{eqnarray}
where [as for all triplets $P_{ij}=\rmi P^r(\sigma ^r)_{ij}$; see
(\ref{RijRr})]
 \begin{eqnarray}
   P^r &\equiv & h^I(\phi) P_{I}^{r}(q)\,, \qquad P_{\tx} ^r \equiv
-\sqrt{\ft{3}{2}} \partial_{\tx} P^r= h^I_{\tx}P_{I}^{r}\,, \nonumber\\
{\cal N}_i^A & \equiv&\frac{\sqrt{6}}{4} f_{iX}^AK^X =\frac{2}{\sqrt{6} }
f_{iX}^A {\cal R}^{r\,YX}  D _Y P^r \,,\qquad K^X\equiv h^I(\phi
)K_I^X(q)\,.
 \label{defcN}
\end{eqnarray}

The scalar potential is given by
 \begin{equation} \label{potential} \cV =
  - 4 P^r P^r +2 P_{\tx}^r P_{\ty}^r
g^{\tx\ty} + 2 {\cal N}_{iA}{\cal N}^{iA} 
\,.
 \end{equation}
This expression can be understood as in all supersymmetric theories
(see~\cite{Cecotti:1986wn} for a proof in 4 dimensions) by squaring the
scalar part of the supersymmetry transformations of the fermions using
their kinetic terms. The kinetic terms of the fermions are
\begin{equation}
  e^{-1}{\cal L}^{\cN = 2}_{\rm ferm,kin}= -\ft12\bar \psi _\mu \gamma ^{\mu \nu \rho }\partial
  _\nu \psi _\rho -\ft12\bar \lambda ^i_{\tx} \gamma ^\nu \partial_\nu  \lambda _i^{\tx}-\bar \zeta
  ^A \gamma ^\nu \partial_\nu\zeta _A\,.
 \label{Lfermkin}
\end{equation}
This defines the metric to be used to square the supersymmetry
transformations:
\begin{equation}
-\ft12(\delta _{\epsilon _1,sc}\bar \psi _\mu) \gamma ^{\mu \nu \rho }
(\delta _{\epsilon _2,sc}\psi _\rho) -\ft12(\delta _{\epsilon _1,sc}\bar
\lambda ^i_{\tx})\gamma ^\nu (\delta _{\epsilon _2,sc} \lambda
_i^{\tx})-(\delta _{\epsilon _1,sc}\bar \zeta^A)\gamma ^\nu(\delta
_{\epsilon _2,sc}\zeta _A)=\ft14g^2\bar \epsilon _1\gamma ^\nu  \epsilon
_2\, {\cal V}\,.
 \label{sqsusyV}
\end{equation}
The gravitino gives the negative contribution to the potential, while the
gauginos and hyperinos give the positive contributions.

We introduce the scalar ``superpotential'' function $W$, which can be read
off the gravitino supersymmetry transformation, by\footnote{As a
convention, we pick a positive definite $W$.}
\begin{equation}
  W=\sqrt{\ft13 P_{ij}P^{ij}}=\sqrt{\ft23 P^rP^r}\,,
 \label{Wmydef}
\end{equation}
such that the potential gets, under certain conditions, the form that has
been put forward for gravitational stability:
\begin{equation}
 {\cal V}= - 6 W^2+\dfrac92 g^{\Lambda\Sigma}\partial_\Lambda
 W\partial_\Sigma W\,, \label{pot}
\end{equation}
where $g_{\Lambda \Sigma}$ is the metric of the complete scalar manifold,
involving the scalars of vector multiplets as well of hypermultiplets.  It
is easy to see that in this case critical points of $W$ are also critical
points of $\cV$.

For one scalar, a proof of gravitational stability was found
in~\cite{Boucher:1984yx} (even without supersymmetry) in 4 dimensions,
and extended to higher dimensions and to the multiscalar case
in~\cite{Townsend:1984iu} for potentials that are a function of the
``superpotential'' as in (\ref{pot}). However, more general potentials are
also compatible with the gravitational stability. More recently, this
issue has been revived in \cite{Skenderis:1999mm,Chamblin:1999cj}.

The negative part of the potential (\ref{potential}) straightforwardly
takes the form of the first term in (\ref{pot}). For the contribution of
the hypermultiplets, the form $g^{XY}\partial_X W\partial_Y W$ follows
from
\begin{equation}
  \partial_X W=\frac{2}{3W}P^rD_XP_r=\frac{2}{3W}P^r {\cal R}^r_{XY}K^Y
 \label{dXW}
\end{equation}
and (\ref{algRSU2}). However, for the vector multiplets the analogous
expression cannot be obtained in general. Using  the decomposition of the
vector $P^r$ in its norm and phases:
\begin{equation}
  P^r=\sqrt{\ft32} W Q^r\,,\qquad Q^rQ^r=1\,,
 \label{PWQ}
\end{equation}
one sees that the term $2P^r_\tx P^{r\tx}$ in (\ref{potential}) gets the
form of the vector multiplet contribution in (\ref{pot}) if
\begin{equation}
  (\partial _\tx Q^r)(\partial ^\tx Q^r)=0\ \Rightarrow\ \partial _\tx Q^r=0\,.
 \label{dxQdxQ}
\end{equation}
This condition\footnote{This constraint is equivalent to the one found
in~\cite{DW}} is satisfied in several cases. When there are no
hyperscalars and only Abelian vector multiplets, the constraints
(\ref{defprep}) and (\ref{constr2prep}) imply that the $Q^r$ are
constants. Also, when there are no physical vector multiplets, this
condition is obviously satisfied. We will see below that (\ref{dxQdxQ})
is related to a condition of unbroken supersymmetry. In the explicit
example that we will show in section~\ref{ss:GaugingFlow}, there will be
flows where $Q^r$ is independent of the scalars in vector multiplets, such
that again (\ref{dxQdxQ}) is satisfied.

\subsection{BPS equations in a domain wall background}\label{ss:DWbg}

We are looking for supersymmetric domain wall solutions that preserve half
of the original supersymmetries of the ${\cal N} =2$ supergravity. Thus
we use as a generic ansatz for the metric
\begin{equation}
\rmd s^2 = a(x^5)^2 \rmd x^{\underline{\mu}} \rmd x^{\underline{\nu}}
\eta_{\underline{\mu\nu}}  + (\rmd x^5)^2\,, \label{metric}
\end{equation}
where $\underline{\mu },\underline{\nu }=0,1,2,3$, which respects
four-dimensional Poincar{\'e} invariance, and we model this solution by
allowing the scalars to vary along the fifth direction $x^5$. These
solutions are obtained when we require that the supersymmetry
transformation rules on this background vanish for some Killing spinor
parameter $\epsilon^i$.

When all vectors are vanishing, the relevant supersymmetry flow equations
for the gravitinos $\psi^i$, the gaugini $\lambda_i^{\tx}$, and the
hyperini $\zeta^A$ are
\begin{eqnarray}
\delta_\epsilon \psi _{\underline{\mu}i} &=&\partial
_{\underline{\mu}} \epsilon _i + \gamma_{\underline{\mu}}
\left(\frac{1}{2} \frac{a'}{a}\gamma _5\epsilon _i +
\frac{\rmi}{\sqrt{6}} g
P_{ij}\epsilon ^j\right)\,,\nonumber\\
\delta_\epsilon \psi _{5i} &=& \epsilon_i ^\prime - q^{X\prime}
\omega_{Xi}{}^{j}\epsilon_j+\frac{\rmi}{\sqrt{6}} g \gamma_5
P_{ij}\epsilon^j \,,\nonumber\\
\delta_\epsilon \lambda_i^{\tx} &=& - \frac{\rmi}{2}\gamma^5
\epsilon_i
\, \phi^{\tx \, \prime} + g P^{\tx}_{ij}\epsilon^j\,, \nonumber\\
  \delta_\epsilon \zeta^A &=&f_{iX}^A\left[ - \ft{\rmi}{2}\gamma ^5
  q^{X\prime} - \frac{2g}{\sqrt{6} }  {\cal R}^{r\,XY}  (D _Y  P)^r\right] \epsilon ^i\,,
 \label{transfos}
\end{eqnarray}
where the prime is a derivative with respect to $x^5$, and we have
assumed that $q^X$ depends only on $x^5$.

\bigskip

The equation $\delta\psi_5^i=0$ gives just the dependence of the Killing
spinor on the fifth coordinate. We assume\footnote{In some cases there
may be other solutions. At the critical points, the supersymmetry is
doubled, the extra Killing spinors being of the type with extra
dependence on the $x^{\underline{\mu }}$. Here we restrict ourselves to
solutions with Killing spinors that do not depend on
$x^{\underline{\mu}}$.} also that the Killing spinor does not depend on
$x^{\underline{\mu}}$.

\bigskip

The first Killing equation  gives
\begin{equation}
\label{Killingpsi4} \rmi \frac{a'}{a}\gamma_5\epsilon_i =
\sqrt{\frac{2}{3}} g P_{ij}\epsilon^j\,,
\end{equation}
whose consistency as a projector equation requires
\begin{equation}
  \left[ \delta _i^k\left( \frac{a'}{a}\right) ^2-g^2\frac{2}{3}
P_{ij}P^{jk}\right] \epsilon_k=0\,.
 \label{consistKilpsi4}
\end{equation}
This can then be easily written in terms of $W$ as
\begin{equation}
\left[ \left( \frac{a'}{a}\right)^2-g^2 W^2\right] \epsilon_i = 0\,.
\label{consistKilpsiSimpl}
\end{equation}
For any preserved supersymmetry, this gives us an equation relating the
warp factor and the superpotential (with $g>0$):
\begin{equation}
g W=\left|\frac{a'}{a}\right|=\pm \frac{a'}{a}\,.
 \label{consistgW}
\end{equation}
Using the notation (\ref{PWQ}), the projection (\ref{Killingpsi4}) is [we
further keep consistently the upper and lower signs as they appear in
(\ref{consistgW})]
\begin{equation}
  \gamma _5\epsilon _i=\pm Q^r\sigma ^r_{ij}\epsilon ^j\,.
 \label{projQ}
\end{equation}

\bigskip

The gaugino equation, after using (\ref{Killingpsi4}), gives rise to the
condition
 \begin{equation} P_{ij} \epsilon^j \phi^{\tx \prime} = - 3
\frac{a^\prime}{a} \, g^{\tx \ty}
\partial_{\ty} P_{ij} \epsilon^j\,.
\label{Gauginotransf}
 \end{equation}
Using the decomposition of $P^r$ in (\ref{PWQ}) one finds
\begin{equation}
W Q^r \phi^{\tx \prime} = - 3 \frac{a^\prime}{a} \, g^{\tx \ty} (Q^r
\partial_{\ty} W + W \partial_{\ty} Q^r).
\end{equation}
Since $Q^r \partial_{\tx}  Q^r=0$, the two  pieces on the right are
orthogonal to each other and so we derive as independent conditions
\begin{equation}
\partial_{\ty} Q^r=0 \label{dQr0}
\end{equation}
and [using (\ref{consistgW})]
\begin{equation}
\phi^{\tx \,\prime} = \mp 3 g\, g^{\tx \ty} \partial_{\ty} W\,.
\label{dWvector}
\end{equation}
The first condition is (\ref{dxQdxQ}), and we thus find that the BPS
condition is equivalent to requiring that the potential can be written in
the stability form (\ref{pot}). Notice that the projection given in
(\ref{projQ}) therefore only depends on the hypermultiplets.\footnote{In
the presence of tensor multiplets, the gaugino supersymmetry (SUSY) rule
would have been modified by an additional term $\delta_\epsilon^\prime
\lambda_i^{\tx} = g W^\tx \epsilon_i$. However, this would have been put
to zero by the gaugino projector equation ${\cal
A}^\tx{}_k{}^i\epsilon_i=({\cal A}^{0\tx} \epsilon_k +{\cal
A}^{r\tx}\rmi(\sigma_r)_k{}^i)\epsilon_i=0$ where ${\cal A}^{0\tx}\equiv
W^\tx$ and ${\cal A}^{r\tx}$ was implicit in (\ref{Gauginotransf}).}

The formula (\ref{dWvector}) can be generalized to the hypermultiplets. In
view of this,  we turn to the hyperino Killing equation, the last of
(\ref{transfos}). For the first term, we can already use (\ref{projQ}).
We multiply the transformation of the hyperinos by $f^Y_{Aj}$. Equations
(\ref{defgXY}) and (\ref{RSU2f}) lead to
\begin{equation}
f_{YjA}f^{iA}_X= \ft12g_{YX}\delta _j{}^i+{\cal R}_{YXj}{}^i\,.
 \label{ffgR}
\end{equation}
This gives
\begin{equation}
0    = \left[g_{YX}\delta _j{}^i
   +2\rmi{\cal R}_{YX}{}^r \sigma ^r{}_j{}^i\right]
\left[   \pm\frac{1}{2} \rmi Q^s\sigma ^s{}_i{}^k q^{X\prime} +
\frac{g\sqrt{6} }{4} K^X\delta _i{}^k  \right] \epsilon _k \,.
 \label{Killing2}
\end{equation}
This we write as a matrix equation ${\cal A}_{Yj}{}^k\epsilon_k=0$:
\begin{eqnarray}
0&=& \left[ {\cal A}^0_Y \delta_ j{}^k +{\cal A}_Y^r\rmi(\sigma
^r)_j{}^k\right] \epsilon _k\,,
\nonumber\\
{\cal A}^0_Y & \equiv  &\frac{g\sqrt{6}}{4}K_Y\mp{\cal R}_{YX}^r Q^r
q^{X\prime}\,,\nonumber\\
{\cal A}_Y^r&\equiv & \pm\frac{1}{2} Q^r q_Y^{\prime}  \mp \varepsilon
^{rst} {\cal R}_{YX}^s  Q^t q^{X\prime} +\frac{g\sqrt{6} }{2}{\cal
R}_{YX}^r  K^X\,. \label{238}
\end{eqnarray}
The reality of these quantities implies that the determinant of the
matrix ${\cal A}_Y$ is
\begin{equation}
  \hbox{det }\cA_Y = \left( {\cal A}^0_Y \right) ^2+\left({\cal A}_Y^r\right) ^2\,.
 \label{detforeps}
\end{equation}
If there are any preserved supersymmetries, then this determinant has to
be zero. Therefore, ${\cal A}^0_Y={\cal A}_Y^r=0$. However, it is easy to
see that the condition ${\cal A}^0_Y=0$ implies ${\cal A}_Y^r=0$, and is
thus the remaining necessary and sufficient condition. With (\ref{dXW}),
this implies
\begin{equation}
 g \partial _X W=\sqrt{\ft23}gQ^r {\cal R}^r_{XY}K^Y= \mp\ft{1}{3}
g_{XY}q^{Y\prime}\,.
 \label{partialXWsol}
\end{equation}
One can also show that this equation is sufficient for the Killing
equations.

We have obtained the same condition for the scalars of hypermultiplets as
for vector multiplets, and we can write collectively for all the scalar
fields $\phi^\Lambda=\{\phi^{\tx},q^X\}$
\begin{equation}
  \phi ^{\Lambda '}=\mp 3 g\, g^{\Lambda \Sigma }\partial _\Sigma W\,.
 \label{qprimeall}
\end{equation}
This equation, together with the constraint (\ref{dQr0}) and the flow
equation for the warp factor (\ref{consistgW}), completely describes our
supersymmetric flow.\footnote{This flow equation also appears
in~\cite{DW}. However, there it was derived using a condition that is
stronger than the one that we need. Our condition is the one that also
implies the stability form of the potential.} A solution to these
equations is also a solution of the full set of equations of motion.

\bigskip

The fact that BPS states are described by equations (\ref{qprimeall}) and
(\ref{consistgW}) can also be seen from the expression of the energy
functional. Once the (\ref{dQr0}) condition is satisfied, such a
functional can be written as
\begin{eqnarray} E &=&  \int_{-\infty}^{+\infty} d x^5
\, a^{4}\left[ {1\over 2}\left(
\phi^{\Lambda '}\mp 3g \partial^\Lambda W\right)^2 - 6 \left(\frac{a^\prime}{a} \mp g W\right)^2 \right] \nonumber \\
&&
 \mp 3g \int_{-\infty}^{+\infty} d x^5 \frac{\partial
\phantom{x^5}}{\partial x^5} \left(a^4 W\right) +
4\int_{-\infty}^{+\infty} d x^5 \frac{\partial \phantom{x^5}}{\partial
x^5} \left(a^3 a^\prime\right)\,. \label{energy}
\end{eqnarray}

\subsection{Renormalization group flow}

The above formulas can be used to obtain the equations that give the
dependence of the scalars on the warp factor $a$. Using the chain rule,
the relevant supersymmetry flow equations for all the scalar fields
reduce to
\begin{equation} \label{betafun}
 \beta^\Lambda \equiv a \frac{\partial\phantom{a}}{\partial a} \phi^\Lambda
= a \frac{\partial x^5}{\partial a} \frac{\partial \phi ^\Lambda
}{\partial x^5} =\mp3g \frac{a}{a^\prime}g^{\Lambda\Sigma}
\partial_\Sigma   W
  = - 3 g^{\Lambda\Sigma} \frac{\partial_\Sigma W}{W} \,.
\end{equation}
The notation as a beta function follows from the interpretation as a
conformal field theory, where the scalars play the role of coupling
constants and the warp factor $a$ is playing the role of an energy scale.

This same function can be used to determine the nature of the critical
points $\phi^*$. Whether $\phi^*$ has to be interpreted as UV or IR in
the dual CFT can be inferred from the expansion of (\ref{betafun})
\begin{equation} \label{2ndeq}
 a \frac{\partial\phantom{a}}{\partial a} \phi^\Lambda  =
\left(\phi^\Sigma   - \phi^{\Sigma  \,
    *}\right) \left. \frac{\partial \beta^\Lambda }{\partial \phi^\Sigma  }\right|_{\phi^*}.
\end{equation}
This tells us that any time the matrix
\begin{equation}
\cU_\Sigma {}^\Lambda  \equiv - \, \left. \frac{\partial \beta^\Lambda
}{\partial \phi^\Sigma }\right|_{\phi^*} =\left. \frac{3}{W}\, g^{\Lambda
\Xi}\frac{\partial^2 W}{\partial \phi^\Sigma \partial \phi ^\Xi
}\right|_{\phi^*}
 \label{defcU}
\end{equation}
has a positive eigenvalue $\phi^*$ is a UV critical point, whereas when
it has negative eigenvalues $\phi^*$ is IR.

The eigenvalues of $\cU$ are the conformal weights $E_0$ of the
associated operators in the conformal picture. One can obtain a general
formula~\cite{Alekseevsky:2001if} for ${\cal U}$:
\begin{equation}
  {\cal U}=\pmatrix{\frac32\delta _X{}^Y- \frac{1}{W^2}{\cal J}_X{}^Z{\cal L}_Z{}^Y&
  \frac1{W^2} {\cal J}_{XZ}\partial ^yK^Z\cr  -\frac1{W^2} (\partial _xK^Z){\cal J}_Z{}^Y
  &2\delta _x{}^y}\,,
 \label{calUres}
\end{equation}
where the first entry corresponds to hypermultiplets and the second to
vector multiplets. The quantities ${\cal J}$ and ${\cal L}$ are defined
as\footnote{This splitting was also put in evidence in
\cite{D'Auria:2001kv}.}
\begin{equation}
 D_X K_Y={\cal J}_{XY}+{\cal L}_{XY} \,, \qquad
 {\cal J}_{XY}=2 P^r{\cal R}^r_{XY} \,.
 \label{defcalJL}
\end{equation}
They commute, ${\cal J}^2$ is proportional to (minus) the unit matrix,
and the trace of ${\cal J}{\cal L}$ is zero:
\begin{equation}
  {\cal J}_X{}^Y{\cal J}_Y{}^Z=-\ft32W^2\delta _X{}^Z\,, \qquad {\cal J}_X{}^Y{\cal
   L}_Y{}^Z={\cal L}_X{}^Y{\cal J}_Y{}^Z\,,\qquad {\cal J}_X{}^Y{\cal
   L}_Y{}^X=0\,.
 \label{propcalJL}
\end{equation}
The decomposition of $DK$ in (\ref{defcalJL}) is a split of the
isometries in SU(2) and USp(2$r$) parts.

The lower right entry of (\ref{calUres}), a consequence of the basic
equations of very special geometry, is the statement that for only vector
multiplets there are only UV critical points, preventing the RS
scenarios~\cite{KL,BC}. The other entries imply that the appearance of IR
directions can be due to two different mechanisms~\cite{DW}. One is the
presence of the hypermultiplets, if the upper left entry gets negative
values, whereas the other is due to the possibility of mixing between
vector and hypermultiplets. To have negative eigenvalues due to the
hypermultiplets only, the ${\cal L}$ matrix has to get large. This means
that the gauging has to be ``mainly'' outside the SU(2) group. We will see
this explicitly in the examples of section~\ref{ss:GaugingFlow}, where
the orthogonal part to SU(2) is a U(1) group.

An immediate consequence of (\ref{calUres}) is that
\begin{equation}
  \trace\, {\cal U}=6\,r + 2\,n \,.
 \label{sumEigen}
\end{equation}
The right-hand side is thus the sum of all the eigenvalues. This implies
that there are no pure IR fixed points, i.e., there are at most fixed
points for which flows in particular directions are of the IR type.

These same eigenvalues are related to the scalar masses through the mass
matrix \cite{FGPW}
\begin{equation}
    {\cM}_{\Lambda}{}^\Sigma = W_{\rm cr}^2 \; \cU_{\Lambda}{}^\Delta
    \, ({\cU_{\Delta}}^\Sigma - 4 {\delta_\Delta}^\Sigma)\,.
    \label{eq:massmatr}
\end{equation}
The scaling dimensions of the dual conformal fields are therefore the
eigenvalues of $\cU$.

Equations (\ref{consistgW}) and (\ref{qprimeall}) also lead directly to
the monotonicity theorem for $A^\prime$. Indeed, defining
\begin{equation}
  A=\ln a \,,\qquad  A'=\pm gW\,,
 \label{defAA}
\end{equation}
we have directly that
\begin{equation}
A''=\pm g W'=-3g^2  (\partial _\Lambda W)g^{\Lambda \Sigma }(\partial
_\Sigma W)\leq 0\,.
 \label{sugractheorem}
\end{equation}
Therefore $A'$ is a monotonically decreasing function. In the usual
holographic correspondence, this is related to the monotonicity of the
$c$ function~\cite{GPPZ2,FGPW,Skenderis:1999mm,Anselmi,Girardello:1999bd}.

\medskip

The above issues can be applied to address the question of the existence
of \textit{smooth Randall--Sundrum} scenarios. In such a scenario, the
scalars should get to a constant value at $x^5=\pm\infty $, and with
(\ref{qprimeall}) this means that $W$ should have an extremum at
$x^5=\pm\infty $, i.e., with (\ref{betafun}), a zero of the beta function
and thus a critical point. For a RS scenario the warp factor should be
small there, i.e., it should be a critical point for a small energy
scale, an IR critical point. Thus, we need a solution that interpolates
between \textit{two IR critical points} for $x^5=\pm\infty $, getting to a
maximum of the warp factor $A$ at the center of the domain wall, placed,
for instance, at $x^5=0$. This requires that at the same point \textit{$W$
should be zero.}

This situation can in principle be realized without violating the
condition (\ref{sugractheorem}). Indeed, take a $W$ that decreases to
zero at $x_5=0$ from positive $x_5$. With a smooth flow, one might expect
that $W$ changes sign, as its derivative is nonzero at this point. Note,
however, that our $W$ is always positive due to its definition as the norm
of the SU(2) vector $P^r$. This is necessary because in the geometry of
hypermultiplets the local SU(2) is essential, and $W$ has to be an
invariant function. It thus bumps up again and increases. But at the same
time the unit vector $Q^r$ jumps to its negative. In this way, $P^r=WQ^r$
behaves smoothly, leading to a smooth flow despite the apparent jumps.
Because of the sign switches in $\partial W$ and $Q$, for negative $x_5$,
one must take consistently all the different signs in equations
(\ref{consistgW})--(\ref{qprimeall}). Note that the two sign flips
combine such that the projection of the Killing spinor in (\ref{projQ})
will not change. Then $W$ will increase again for negative $x_5$ and the
monotonicity of the warp factor will not be violated.

Of course in the holographic interpretation of the $c$ theorem, the
central charge would blow up or the height function would become singular
at the zero of the superpotential~\cite{Behrndt:2001qa}, and the dual
field theory would be ill defined at that point. In spite of this, the
supergravity monotonicity theorem can further be satisfied with
increasing $W$, if it was decreasing at the other side of $x^5=0$.

The interesting points are thus the zeros that we just discussed, and the
critical points,  where $\partial _\Lambda W=0$. We now turn to
discussing the properties of the latter.

\subsection{Enhancement of unbroken supersymmetry and \\
algebraic attractor equations}

The search for critical points can be nicely formalized as an attractor
mechanism, which was discovered in~\cite{attractflows} and was studied in
great detail in the absence of hypermultiplets. So far only partial
investigations exist for the coupling of both vector multiplets and
hypermultiplets, in the context of domain
walls~\cite{DW,Behrndt:2001qa,Behrndt:2001mx} and in the context of the
BPS instantons~\cite{Gutperle:2000ve}. At the fixed points of the
solution the moduli are defined by the condition $\cN_{iA} = 0$
\cite{AnnaGianguido}, which implies $K^X=0$, as can be understood from
(\ref{defcN}) and (\ref{dXW}). This fact was also observed
in~\cite{Gutperle:2000ve,DW,Behrndt:2001qa,D'Auria:2001kv}.

Here we will use the fact derived in the previous section, that the
Killing spinor projector $Q^r$ must satisfy (\ref{dxQdxQ}). Only in such
case does the superpotential $W$ control the flow equations. {\it Using
the enhancement of supersymmetry near the critical points we will derive
all necessary and sufficient conditions for critical points.} Our method
follows~\cite{attractflows,Chamseddine:1997pi,Chou:1997ba}, where the
geometric tools of special geometry in $D=4$ and very special geometry in
$D=5$ were used to convert the BPS differential equations into algebraic
ones and where enhancement of unbroken supersymmetry played an important
role.

Consider the domain wall solutions of the previous subsection in the
limit where the scalars are frozen:
\begin{equation}
q^{X'}=0 \ , \qquad  \phi^{x'}=0 \ , \qquad {a'\over a}=\pm W_{\rm
cr}=\rm {const} \,. \label{ads}
\end{equation}
If const $\neq 0$, the flow  tends to the AdS horizon in case of the IR
critical point and to the boundary of the AdS space in case of the UV
critical point. This becomes clear when the metric is rewritten as $\rmd
s^2= a^2 (\rmd x^{\underline{\mu}})^2 + {1\over W^2}\left( \rmd a\over
a\right)^2$. For constant {\it nonvanishing $W$}, small $a$ define the
horizon of the AdS space  whereas large $a$ correspond to its boundary.

The gravitino supersymmetry transformation at the critical point
(\ref{ads}) acquires a second Killing spinor. This is the same doubling
that always occurs in the AdS background near the black hole horizon. One
finds that
\begin{equation}
\delta_{\epsilon} \psi_{\mu}=0\,, \qquad (\epsilon_i)_{\rm attr}  \neq
0\,, \label{Crit1}
 \end{equation}
without  restrictions on the Killing spinors, i.e., they have 8 real
components.

In analyzing the equations we will have to be careful that we are inside
the domain of validity of our coordinate system. In particular, this means
that $g_{xy}$, $g_{XY}$, $f^{iA}_X$, and ${\cal R}^{r\,XY}$ are neither
vanishing nor infinite. We will be able to invert these geometric objects
using the rules of very special and quaternionic geometry. The procedure
is analogous to the steps performed in the previous section to find the
solutions with $\cN = 1$ unbroken supersymmetry. Now we will specify it
to the case of frozen moduli and $\cN = 2$ unbroken supersymmetry.

By direct inspection of the supersymmetry transformations we observe that
the first term in the gaugino and hyperino transformation vanishes and we
get
\begin{eqnarray}
\delta_\epsilon \lambda_i^{\tx} &=&  g P^{\tx}_{ij}\epsilon^j=0\,, \nonumber\\
  \delta_\epsilon \zeta^A &=&f_{iX}^A\left[  \frac{\sqrt{6}g }{4}  K^X\right] \epsilon
  ^i=0\,.
 \label{enhan}
\end{eqnarray}
The first of these equations for $\epsilon_i\neq 0$ can be satisfied if
and only if
\begin{equation}
(\partial_{\ty} P^{r})_{\rm attr}= 0\,,\label{Crit2}
\end{equation}
which, for AdS vacua, can also be written as
\begin{equation}
(Q^r
\partial_{\ty} W + W \partial_{\ty} Q^r)=0\quad \Rightarrow\quad
(\partial_{\ty} W)_{\rm attr} =0 \quad \mbox{and}\quad   (\partial_{\ty}
Q^r)_{\rm attr}=0\,. \label{eq:Crit2b}
 \end{equation}
The implication follows from the same argument as for (\ref{dQr0}).

Finally, we have to derive the necessary and sufficient conditions to
satisfy the hyperino equation also. Evaluating (\ref{238}) at the
attractor point $q^{\prime X} = 0$ for $\epsilon_i\neq 0$ requires
\begin{eqnarray}
({\cal A}^0_Y)_{\rm attr} \equiv  \frac{g\sqrt{6}}{4}K_Y =0\,, \qquad
({\cal A}_Y^r)_{\rm attr} \equiv  \frac{g\sqrt{6} }{2}{\cal R}_{YX}^r
K^X=0\,.
 \label{hypCrit}
\end{eqnarray}
The solution of this equation is
\begin{equation}
 ( K^X )_{\rm attr} \equiv h^I(\phi) K_I^X ( q)=0\,.
 \label{Crit3}
\end{equation}
As previously noticed, this is an algebraic equation that defines the
fixed values of the scalars at the critical point and solves
$\delta_\epsilon \zeta^A=0$.

The algebraic rather than differential nature of this condition
stimulates us to look for an algebraic equation also in the vector
multiplet sector of the theory. Indeed, such an algebraic attractor
equation was known to be valid for AdS critical points in theories without
hypermultiplets \cite{Chou:1997ba} and we now try to generalize it. We
start with (\ref{Crit2}) and multiply this equation by $g^{\tx \ty}
\partial_{\ty}  h_I$. Using the fact that $(h_{Ix},h_I)$ forms an
$(n+1)\times (n+1)$ invertible matrix in very special geometry
(\ref{defag}), this equation becomes
\begin{equation}
P^r_I = C_{IJK}h^J h^K P^r =h_I P^r\,. \label{GauginoCrit1}
 \end{equation}

So far the result is valid for any critical point. If we now restrict
ourselves to the AdS ones, we can multiply (\ref{GauginoCrit1}) by $Q^r$
and get
\begin{equation}
h_I W = C_{IJK}\tilde h^I \tilde h^K  = P_I\,,\qquad P_I(q)\equiv\sqrt
{\ft23} P^r_I Q^r\,, \qquad \tilde h^I(\phi ,q) =h^I \sqrt W\,.
\label{algebraicV}
\end{equation}
Note that $P_I$ depends only on quaternions. This type of algebraic
attractor equation with constant values of $P_I$ was used in an efficient
way in various situations before. In particular, it was used in
calculations of the entropy of Calabi--Yau black holes and the warp
factor of Calabi--Yau domain walls near the critical points. We have
shown here that in the presence of the hypermultiplets the analogous
algebraic equations with quaternion-dependent $P_I$ are valid at the
critical points. Thus the algebraic equation $
 h_I P^r  = P_I^r\,
$ is equivalent to the differential equation $\partial_x W=0$, since for
supersymmetric flows the $\partial_x Q^r=0$ condition is satisfied. If we
multiply this by $h^I$ we will get an identity $P^r = P^r$; however,
(\ref{GauginoCrit1}) is not satisfied in general but only at the fixed
points where all scalars are constant.

Thus we have the system of algebraic equations, defining the critical
points:
\begin{equation}
 h_I(\phi_{\rm cr}) P^r (\phi_{\rm cr}, q_{\rm cr}) = P_I^r( q_{\rm cr}) ,
 \qquad  K^X (\phi_{\rm cr}, q_{\rm cr}) =0\,.
\label{algebraicV+H}
\end{equation}
They are equivalent to the system of differential equations minimizing the
superpotential. These equations, together with $\partial_x Q^r=0$, are
equivalent to minimizing the triplet of the prepotentials
\begin{equation}
 (\partial_{\tx} P^{r})_{\rm cr}=0, \qquad  (D_X P^r)_{\rm cr}=0 \,.
\label{diftriplet}
\end{equation}
The  advantage of having  algebraic rather than differential equations
defining the critical points is already obvious in the simple examples
that we consider in the models below, but it will be even more essential
in cases with arbitrarily many moduli.

As a conclusion of this section, we can summarize the relevant equations
to be examined in the specific examples. Given a scalar manifold, we have
to look for the following special points.
\begin{enumerate}
  \item \textit{Fixed points}. These are points where $\partial_\Lambda W=0$.
  They are determined by algebraic equations:
\begin{equation}
{\rm{Fixed\ points:}} \qquad K^X\equiv h^I(\phi)K^X_I(q)=0\,,\qquad
P_I^r(q) = h_I(\phi) P^r(\phi,q)\, . \label{fixedpoints}
\end{equation}
In particular, for the AdS case, the eigenvalues of the matrix
(\ref{defcU}) determine whether they are UV
  (eigenvalues positive), or whether some eigenvalues are negative. In
  the latter case, they can be used as IR fixed points, and represent the
  values of the scalars at $x^5=\pm\infty $ in the RS scenario.
  \item \textit{Zeros}. These determine the values of the scalars on the
  place of the domain wall, i.e., where the warp factor reaches an
  extremum:
\begin{equation}
{\rm{Zeros:}} \qquad   P^r=h^I(\phi )P^r_I(q)=0\,.
 \label{zeros}
\end{equation}
Note that the presence of zeros implies that the $\beta $ function
diverges. This indicates that the AdS/CFT correspondence breaks down at
this point. These zeros are necessary for Randall--Sundrum domain walls
but are thus pathological for applications as renormalization group flows.
\end{enumerate}

\section{A model with a vector and a hypermultiplet}\label{ss:scManIsom}
In this and the next section we want to specify the results obtained so
far to two detailed examples. The simplest model that can show all the
main features of this kind of analysis is given by supergravity coupled
to one vector and one hypermultiplet. Thus, as a first step, we describe
in full detail the toy model based on the universal hypermultiplet alone.
Then we analyse the complete model, whose moduli space is given by the
scalar manifold
\begin{equation}
    {\cal M} = \textrm{O(1,1)} \times \frac{\textrm{SU(2,1)}}{\textrm{SU(2)}\times \textrm{U(1)}}\,.
    \label{scalarM}
\end{equation}
Actually, in this example we focus on one important supersymmetric domain
wall solution that was previously discussed as the dual to the
renormalization group flow describing the deformation from an ${\cal N} =
4$ to an ${\cal N}=1$ super Yang--Mills theory with SU(2) flavour
group~\cite{FGPW}.

This solution (at least numerically) was originally obtained inside the
${\cal N}=8$ gauged supergravity theory, but we will show that it can also
have a consistent description in the standard ${\cal N} = 2$ one. Notice
that this same flow was claimed to be present also in a truncated ${\cal
N} = 4$ gauged supergravity coupled to two tensor multiplets, but the
relevant model has only recently been constructed
in~\cite{Dall'Agata:2001vb}. The older reference~\cite{N4stand}  dealt
only with the coupling to vector multiplets and gauging of the SU(2)${}_R$
group, whereas~\cite{N4Rom,LPT} discussed the $\textrm{SU(2)}_R\times
\textrm{U(1)}_R$ gauging without any matter coupling.

\medskip

More precisely,  we will show that {\it  the FGPW flow can be
consistently retrieved in the ${\cal N} = 2$ supergravity with one
massless graviton multiplet, a massless vector multiplet, and one
hypermultiplet with the gauging of a $\textrm{U(1)} \times \textrm{U(1)}$
symmetry of the scalar manifold (\ref{scalarM}).}

\medskip

In the decomposition~\cite{FGPW} of the $\cN=8$ graviton multiplet into
$\cN=2$ multiplets (which is completely valid only at the infrared fixed
point), the supergravity fields are arranged into representations of the
$\textrm{SU}(2,2|1)\times \textrm{SU(2)}_I$ residual superalgebra.
Retaining only $\textrm{SU(2)}_I$ singlets leaves us with one graviton
multiplet, one hypermultiplet, one massive vector multiplet, and one
massive gravitino multiplet. However~\cite{FGPW}, since the only scalars
that change along the flow are the two belonging to the massive vector
multiplet, it is expected that the theory  can be further consistently
truncated to one containing the graviton and massive vector multiplets
only, but how can we describe such couplings in the standard
framework~\cite{AnnaGianguido}?

The representations of the $\textrm{SU}(2,2|1)$ supergroup not only
include a massless short graviton multiplet and an arbitrary number of
(massless) short vector and (massive) tensor and hypermultiplets, but
also present a wide spectrum of long and semilong supermultiplets. While
in the general theory~\cite{AnnaGianguido} the couplings and interactions
of short multiplets are explicitly described, those of massive vector
multiplets can arise as the result of a Higgs mechanism where a massless
vector eats a scalar coming from a hypermultiplet.

Since the UV fixed point should correspond to the $\cN = 8$
supersymmetric theory, both the graviphoton and gauge vectors must be
massless there, as they are both gauge vectors of $\textrm{U}(1)\times
\textrm{U}(1) \subset \textrm{SU}(4)$. Then, along the flow, only one of
them (or at most a combination of the two) will remain massless, while the
other will gain a mass, breaking the residual invariance to the
$\textrm{U}(1)_R$ subgroup and giving rise to the massive vector
multiplet described above. This means that we can further decompose the
long vector multiplet into a massless one plus a hypermultiplet, which is
exactly the content of the model we are going to analyse now.

To complete the characterization of the flow we only need some
information to sort out which $\textrm{U}(1) \times \textrm{U}(1)$
subgroup of the isometry group of the manifold $\cM$  has to be gauged.
This can be understood by examining the mechanism that gives the mass to
one of the two vectors.

The vector mass terms come from the kinetic terms of the hypermultiplet
scalars. Indeed, due to the gauged covariant derivatives (\ref{calDbos}),
the kinetic term for such scalars is
\begin{equation}
- \ft{1}{2} (\partial_\mu q^X + g A_\mu^I K_I^X)^2\,, \label{massterm}
\end{equation}
and therefore for a U(1) gauging one has a term like $g^2 A_\mu A^{\mu}
K^2$ in the action, where $A_\mu$ is a linear combination of gauge
vectors and $K^X$ is the corresponding Killing vector of the gauged
isometry. This, of course, will act as a mass term for the $A_\mu$ vector
any time the Killing vector has a nonzero norm.

It is therefore quite easy now to identify the isometries to be gauged in
order to obtain the FGPW flow. They are those associated with Killing
vectors $K_I^X$ that have vanishing norm at the UV fixed point, and such
that along the flow the norm of a combination of them still remains zero.

\subsection{The scalar manifold}

We now turn to the description of the parametrization and of the
isometries of the scalar manifold (\ref{scalarM}). The $\textrm{O}(1,1)$
factor is relative to the vector multiplet scalar $\rho$, and is given by
a very special manifold characterized by $n_V+1=2$ functions $h^I(\rho)$
constrained to the surface (\ref{Chhh1}). Its essential geometric
quantities, the $C$ constants that determine the embedding of this
manifold in the ambient space and also fix the Chern--Simons coupling,
can be chosen to be all but $C_{011}$ equal to zero. Then we
take\footnote{The numerical factors are partly chosen  for convenience
and partly to satisfy the request that there exist a point of the
manifold where the metric $a_{IJ}$ can be put in the form of a delta
$\delta_{IJ}$.}~\cite{GunZag}
\begin{eqnarray}
&&C_{011} = \dfrac{\sqrt{3}}{2}\,, \qquad  h^0 = \frac{1}{\sqrt{3}}\,
\rho^4\,, \qquad  h^1 = \sqrt{\frac{2}{3}} \frac{1}{\rho^2}\,, \nonumber\\
&&g_{\rho\rho} = \frac{12}{\rho^2}\,, \qquad  a_{00} =
\frac{1}{\rho^8}\,, \qquad  a_{11} = \rho^4\,, \qquad a_{01} = 0\,.
\end{eqnarray}
The metric $g_{\rho\rho}$ is well behaved for $\rho \neq 0$ and, for
definiteness, we choose the branch $\rho > 0$.

Much more can be said about the second factor of ${\cal M}$, and due to
its fundamental role we would like to describe it in some more detail. It
is known that the quaternionic K{\"a}hler space
$\frac{\mathrm{SU}(2,1)}{\mathrm{SU}(2)\times \mathrm{U}(1)}$, classically
parametrized by the universal hypermultiplet, is also a K{\"a}hler manifold
\cite{FerraraSabharwal}. This means that it can be derived from a K{\"a}hler
potential, which is usually taken to be
\begin{equation}
\cK = -\ft{1}{2} \log (S + \bar S - 2 C\bar C)\,.
\end{equation}

In addition to the many parametrizations existing in the literature, the
one that will prove convenient to us follows very closely the notations of
\cite{Lukas:1998tt}, with some further redefinitions to match the
conventions of \cite{AnnaGianguido}. We thus call the four hyperscalars
$q^X= \{V,\sigma, \theta, \tau\}$, which are related to the previous
variables by
 \begin{equation}
S = V + (\theta^2 + \tau^2) + \rmi \sigma\,,\qquad  C = \theta - \rmi
\tau\,.
\end{equation}
The domain of the manifold is covered by $V>0$. Note that this
parametrization of the universal hypermultiplet is the one that comes out
naturally from the Calabi--Yau compactifications of $M$ theory \cite{CCDF,
Lukas:1998tt} and thus one can hope to explicitly see how gauging of
isometries can be obtained from such a higher-dimensional description.

\bigskip

Let us define the following one--forms:
\begin{equation}
u  = \frac{\rmd \theta + \rmi \rmd  \tau}{\sqrt{V}}\,, \qquad
v  =  \frac{1}{2V} \left[ \rmd V + \rmi \left( \rmd \sigma - 2 \tau \rmd
\theta + 2 \theta \rmd \tau \right)\right]\,, \label{v}
\end{equation}
which will be very useful in the whole construction. The quaternionic
vielbeins $f^{iA}=f^{iA}_X dq^X$ are then chosen to be [$\varepsilon_{12}
= C_{12} = +1$ are the conventions for the $\mathrm{SU}(2) \times
\mathrm{USp}(2)$ metrics]
\begin{equation}
f^{iA} = \left(
\begin{array}{cc}
    u & -v  \\
    \bar v & \bar u
\end{array}\right)\,, \quad
f_{iA} = \left(
\begin{array}{cc}
    \bar u & -\bar v \\
    v & u
\end{array}\right)\,.
    \label{viel}
\end{equation}
The metric is then given by $g = f^{iA} \otimes f_{iA} = 2 u
\otimes \bar u + 2 v \otimes \bar v$ and reads
\begin{equation}
\rmd  s^2 = \frac{\rmd V^2}{2V^2} + \frac{1}{2V^2}\left( \rmd \sigma + 2
\theta \, \rmd  \tau - 2 \tau \, \rmd  \theta\right)^2 + \frac{2}{V} \,
\left(\rmd \tau^2 + \rmd  \theta^2\right) \,. \label{quatmetric}
\end{equation}
The determinant for such a metric is $1/V^6$ and therefore the metric is
positive definite and well behaved for any value of the coordinates
except $V =0$. Since in the Calabi--Yau derivation $V$ acquires the
meaning of the volume of the Calabi--Yau manifold, we restrict it to the
positive branch $V >0$.

{}From the  vielbeins we can  derive the SU(2) curvature:
\begin{equation}
{\cR_i}^j = -\frac{1}{2}
 f_{iA} \wedge f^{jA} =-\frac{1}{2}
 \left(
\begin{array}{cc}
   -(v \bar v +u \bar u) & 2\bar u \bar v \\
    2 vu & (v \bar v +u \bar u)
\end{array}\right)\,.
\label{curv}
\end{equation}
Using the triplet of curvatures as in (\ref{cRXYij}),
\begin{eqnarray}
{\cal R}^1 & = &- \frac{1}{2V^{3/2}} \left[(\rmd \sigma + 2 \theta \, \rmd
\tau) \rmd \, \theta - \rmd \tau \, \rmd V\right]\,, \nonumber\\
{\cal R}^2 & = &- \frac{1}{2V^{3/2}} \, \left[ (\rmd  \sigma - 2 \tau \rmd
\theta) \rmd \tau + \rmd \theta \rmd  V \right]\,, \nonumber\\
{\cal R}^3 & = & -\frac{1}{V}\, \rmd \theta \rmd \tau +\frac{1}{4V^2} \,
\left[ (\rmd \sigma - 2 \tau \rmd \theta + 2 \theta \rmd  \tau)\rmd V
\right]\,. \label{R123}
\end{eqnarray}
These can be derived from the following SU(2) connections:
\begin{equation}
\omega^1  =  -\frac{\rmd \tau}{\sqrt{V}}\,, \qquad \omega^2  = \frac{\rmd
\theta}{\sqrt{V}}\,,\qquad  \omega^3  =   -\frac{1}{4V}\left( \rmd \sigma
- 2 \tau \rmd \theta + 2 \theta \rmd \tau\right)\,. \label{omega123}
\end{equation}
\subsection{The isometries}
The metric (\ref{quatmetric}) has an SU(2,1) isometry group
generated by the following eight Killing vectors $k^X_\alpha$:
\begin{equation}
\begin{array}{l}
\vec{k}_1 = \left(
\begin{array}{c}
0 \\ 1 \\ 0 \\ 0
\end{array}
\right)\,,
\quad
\vec{k}_2 = \left(
\begin{array}{c}
0 \\ 2\theta \\ 0 \\ 1
\end{array}
\right)\,,
\quad
\vec{k}_3 = \left(
\begin{array}{c}
0 \\ -2\tau \\ 1 \\ 0
\end{array}
\right)\,,
\quad
\vec{k}_4 = \left(
\begin{array}{c}
0 \\ 0 \\ -\tau \\ \theta
\end{array}
\right)\,, \\
\vec{k}_{5} = \left(
\begin{array}{c}
V \\ \sigma \\ \theta/2 \\ \tau/2
\end{array}
\right)\,,
\quad
\vec{k}_{6} =\left(
\begin{array}{c}
2V \sigma  \\ \sigma^2 - \left(V + \theta^2 + \tau^2\right)^2  \\
\sigma \theta - \tau \left(V + \theta^2 + \tau^2\right)\\
\sigma \tau + \theta \left(V + \theta^2 + \tau^2\right)
\end{array}
\right)\,,
\\
\vec{k}_{7} =\left(
\begin{array}{c}
- 2V \theta\\ -\sigma\theta + V\tau + \tau\left(\theta^2 + \tau^2\right) \\
\frac{1}{2} \left(V  - \theta^2 + 3\tau^2\right) \\
-2 \theta \tau - \sigma/2
\end{array}
\right)\,,
\quad
\vec{k}_{8} =\left(
\begin{array}{c}
- 2V \tau\\ -\sigma\tau - V\theta - \theta\left(\theta^2 + \tau^2\right)  \\
-2 \theta \tau + \sigma/2\\
\frac{1}{2}\left(  V  + 3\theta^2 - \tau^2\right)
\end{array}
\right)\, .
\end{array}
\label{killvec}
\end{equation}
The first three correspond to some constant shift of the coordinates; in
particular, the first one  was analysed in~\cite{Lukas:1998tt}, where
$\sigma \to \sigma +c$, whereas the second and third   correspond to the
shifts $\theta \to \theta +a $, $\tau \to \tau + b$, and $\sigma \to
\sigma -2a \tau+ 2 b \theta$. The fourth Killing vector is the generator
of the rotation symmetry between the $\theta$ and $\tau$ coordinates:
$\theta \to \cos \phi \, \theta - \sin \phi \, \tau$, $\tau \to \sin \phi
\, \tau + \cos \phi \, \theta$, which is the one considered in~\cite{DW}.
Finally, the fifth Killing vector is the generator of dilatations while
the remaining three are other complicated isometries of the metric.

The commutators of these vectors confirm that they really close the
$\mathrm{SU}(2,1)$ algebra. To this purpose, it is easier to recast them
in the following combinations
\begin{equation}
\begin{array}{lcl}
\begin{array}{rcl}
SU(2) && \left\{
\begin{array}{rcl}
T_1 & = & \ft{1}{4}\left(k_2 -2 k_8\right),   \\[1.5mm]
T_2 & = & \ft{1}{4}\left(k_3 -2 k_7\right),  \\[1.5mm]
T_3 & = & \ft{1}{4}\left(k_1+k_6-3k_4\right),
\end{array}
\right. \\
&& \phantom{AA} \\
U(1) && \left\{ T_8  =  \dfrac{\sqrt{3}}{4} \left( k_4 + k_1+k_6\right),
\right.
\end{array}  & &
\dfrac{SU(2,1)}{U(2)} \left \{
\begin{array}{rcl}
T_4 & = &  \rmi k_5\,, \\[1.5mm]
T_5 & = & - \rmi\ft{1}{2}\left(k_1-k_6\right), \\[1.5mm]
T_6 & = & - \rmi\ft{1}{4}\left(k_3 +2 k_7\right),\\[1.5mm]
T_7 & = & - \rmi\ft{1}{4}\left(k_2 +2 k_8\right).
\end{array} \right.
\end{array}\label{Tgenerators}
\end{equation}
These generators satisfy the SU(3) commutation relations
\begin{equation}
T_\alpha ^Y\partial _Y T_\beta ^X-T_ \beta ^Y\partial _Y T_\alpha^X=  -
f_{\alpha \beta }{}^\gamma  T_\gamma^X  \,.
\end{equation}
The factors of $\rmi$ in (\ref{Tgenerators}) allow us to have completely
antisymmetric structure constants $f_{\alpha \beta \gamma }= f_{\alpha
\beta }{}^\gamma $ with $f_{123}=1$, $f_{147}=1/2$, $f_{156}=-1/2$,
$f_{246}=1/2$, $f_{257}=1/2$, $f_{345}=1/2$, $f_{367}=-1/2$,
$f_{458}=\sqrt{3}/2$, $f_{678}=\sqrt{3}/2$. The generators $T_4$, $T_5$,
$T_6$, and $T_7$ are imaginary, such that the real algebra is
$\mathrm{SU}(2,1)$.

The relation  (\ref{valuePrI}) leads directly to the prepotentials
\begin{equation}
\begin{array}{rcl}
\vec{P} &=& \left(
\begin{array}{c}  0 \\ 0  \\ -\frac{1}{4V} \end{array}
\right)\,,
\quad
\left(
\begin{array}{c} -\frac{1}{\sqrt{V}} \\ 0  \\ -\frac{\theta}{V} \end{array}
\right)\,,
\quad
\left(
\begin{array}{c}  0 \\ \frac{1}{\sqrt{V}}  \\ \frac{\tau}{V}\end{array}
\right)\,,
\quad
\left(
\begin{array}{c}  -\frac{\theta}{\sqrt{V}} \\ -\frac{\tau}{\sqrt{V}} \\
\frac{1}{2} - \frac{\theta^2+ \tau^2}{2V} \end{array}
\right)\,, \\
&&
\left(
\begin{array}{c}  -\frac{\tau}{2\sqrt{V}} \\ \frac{\theta}{2\sqrt{V}}
\\ -\frac{\sigma}{4V} \end{array}
\right)\,,
\quad
\left(
\begin{array}{c} -\frac{1}{\sqrt{V}}\left[\sigma\tau +
\theta\left(-V+\theta^2+ \tau^2\right)\right] \\
\frac{1}{\sqrt{V}}\left[\sigma\theta -
\tau\left(-V+\theta^2+ \tau^2\right)\right]    \\
-\frac{V}{4}-\frac{1}{4V}\left[\sigma^2 + \left(\theta^2+
\tau^2\right)^2\right] + \frac{3}{2}\left(\theta^2+ \tau^2\right)
\end{array} \right)\,,
\\
&&
\left(
\begin{array}{c}  \frac{4\theta\tau + \sigma}{2\sqrt{V}} \\
\frac{3\tau^2- \theta^2}{2\sqrt{V}} - \frac{\sqrt{V}}{2}  \\
-\frac{3}{2}\tau+\frac{1}{2V}\left[\sigma\theta+ \tau \left(\theta^2+
\tau^2\right)\right]\end{array} \right)\,, \quad \left(
\begin{array}{c} -\frac{3 \theta^2-\tau^2}{2\sqrt{V}} + \frac{\sqrt{V}}{2}  \\
     \frac{\sigma-4\theta\tau }{2\sqrt{V}} \\
\frac{3}{2}\theta+\frac{1}{2V}\left[\sigma\tau- \theta \left(\theta^2+
\tau^2\right)\right] \end{array} \right)\,.
\end{array}
\label{prep}
\end{equation}
As we have seen in the previous sections, these prepotentials are indeed
necessary in order to write an explicit expression for the potential
(\ref{potential}) and superpotential (\ref{Wmydef}) of the gauged theory
and thus  to solve the supersymmetry flow equations (\ref{qprimeall}).

\section{Gauging and the flows}\label{ss:GaugingFlow}

In this section, we will analyse the flows that can be obtained in this
model. This requires, as pointed out in section~\ref{ss:flowtools}, that
we perform a specific gauging and search for critical points and for
zeros of the corresponding potential.

\subsection{Toy model with only a hypermultiplet}\label{ss:toymodel}
We now turn to analyzing the supersymmetric flows that can be obtained in
our model with one vector multiplet and one hypermultiplet. To this end,
it is interesting to start with a preliminary study  on the vacua
obtained by considering  only a U(1) gauging of the universal
hypermultiplet manifold, when no other vectors but the graviphoton are
present. This will give us some important hints.

As for this case $h^0=1$ is the only component of $h^I$, the conditions
for  critical points (\ref{fixedpoints}) just reduce to the vanishing of
a certain linear combination $K$ of the Killing vectors that give rise to
the gauged isometry. In other words, a critical point should be left
invariant under the U(1) generated by $K$. Therefore, such U(1) must be
part of the isotropy group of the manifold,\footnote{This important
feature was not realized in any of the revisions
of~\cite{Behrndt:2000ph}, where isometries were gauged outside the
compact subgroup.} i.e.,
\begin{equation}
  \mathrm{U}(1)_{\rm gauge}\subset \mathrm{SU}(2)\times \mathrm{U}(1)\,.
 \label{U1inistropy}
\end{equation}
In a symmetric space any point is equivalent, and for convenience we have
chosen the basis (\ref{Tgenerators}) such that the generators $T_1$,
$T_2$, $T_3$, and $T_8$ are the isotropy group of the point
\begin{equation}
  (V,\sigma ,\theta ,\tau)=(1,0,0,0)\,.
 \label{basepoint}
\end{equation}
The Killing vector that we consider is given by
\begin{equation}
 K=\sum_{r=1}^3 \alpha_rT_r+\beta \frac{1}{\sqrt{3}}T_8\, ,
\end{equation}
and the same constant parameters $\{\alpha_r,\beta\}$ are used to define
the gauged prepotential $P^r$. Correspondingly, we find
\begin{equation}
  \left.W^2\right|_{(1,0,0,0)}= \ft16 \alpha_r\alpha_r \,.
 \label{W2basis}
\end{equation}
The value of $W$ is indeed determined only by the gauging of the SU(2)
part of the isotropy group of the corresponding point. Therefore, any
gauging in the direction of $T_8$ alone can give rise to a supersymmetric
vacuum of the theory corresponding to a Minkowski space. This same
feature can also be observed for the gauging of an $\mathrm{SU}(2) \times
\mathrm{U}(1)$ group in the $\cN = 4$ theory, namely, the gauging of the
U(1) isometry alone gives rise only to Minkowski vacua
\cite{N4Rom,Dall'Agata:2001vb}. This implies that, if a dual field theory
could be built, it would be in a confined phase.

The matrix ${\cal J}$ in (\ref{defcalJL}) is at the base point
(\ref{basepoint}) proportional to $\alpha_r $, while ${\cal L}$ is
proportional to $\beta $, these being the gaugings in the SU(2) and
orthogonal directions, respectively. The eigenvalues of ${\cal J}{\cal L}$
are twice $\pm \alpha \beta /4$ (where $\alpha $ is the length of the
vector $\alpha_r$) such that the matrix of second derivatives
(\ref{defcU}) according to (\ref{calUres}) satisfies
\begin{equation}
\left.  \mathop{\rm eigenvalues\ }\nolimits {\cal
U}_X{}^Y\right|_{(1,0,0,0)}= \frac{3}{2}\left\{ 1+\frac{\beta }{\alpha
},1+\frac{\beta }{\alpha },1-\frac{\beta }{\alpha },1-\frac{\beta }{\alpha
}\right\},
 \label{EigenddW}
\end{equation}
and this tells us the nature of the critical point. As an example,
consider the gauging in the direction of $\alpha =\alpha _3$. Then the
first two eigenvalues correspond to the $(V,\sigma )$ directions, and the
latter two are of the $(\theta ,\tau )$ directions.

In general, the obtained eigenvalues reflect the supersymmetry structure
of the universal hypermultiplet. We know that the above eigenvalues are
related to the masses of the fields and therefore also to their conformal
dimension $E_0$ (actually, given the eigenvalues $\delta_k$, the relation
is $E_0 = |\delta_k|$ or $E_0 = |4-\delta_k|$). For a hypermultiplet two
of the scalars must have the same $E_0$ and two must have $E_0 + 1$. In
this case this is realized by the fact that two have $E_0 =
\frac{3}{2}\pm \frac{3\beta }{2\alpha}$ and two have $E_{0}=
\frac{5}{2}\pm \frac{3\beta }{2\alpha}$.

\begin{figure}
\begin{center}
\leavevmode \epsfxsize=6cm
 \epsfbox{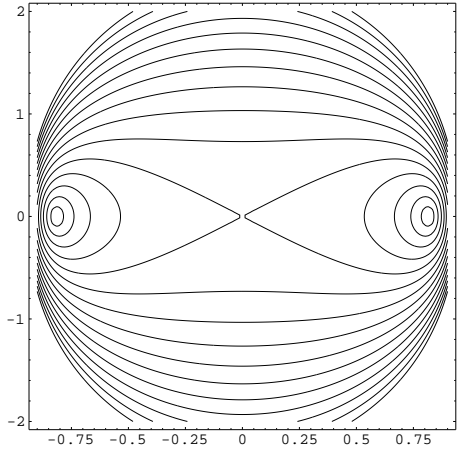}
\caption{\it Contours of constant $W$ in the plane $(\tau ,\sigma )$ with
$V=1 -\tau^2$, $\theta =0$,  for $\alpha_1=\alpha_2=0$,
$\alpha_3=\sqrt{6}$, and $\beta =2\sqrt{6}$. \label{fig:conpl34}}
\end{center}
\end{figure}

\begin{figure}
\unitlength=1mm
\begin{center}
\begin{picture}(120,50)(0,0)
\put(20,0){\leavevmode \epsfxsize=8cm
 \epsfbox{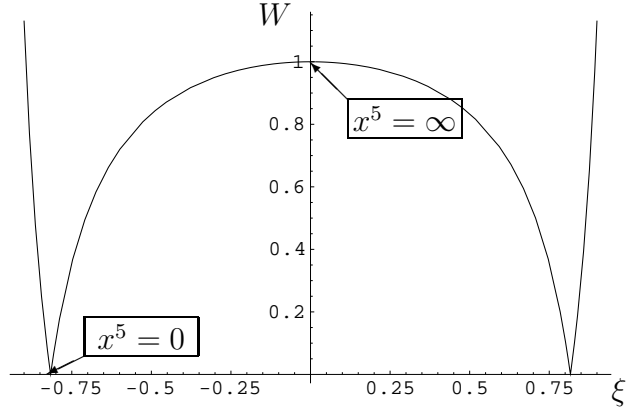}}
\put(65,35){\framebox(15,5){$x^5=\infty $}}
 \put(65,40){\vector(-1,1){5}}
 \put(100,0){$\xi $}
 \put(53,50){$W$}
\put(30,6){\framebox(15,5){$x^5=0 $}}
 \put(30,6){\vector(-2,-1){5}}
\end{picture}
\caption{\it $W$ as a function of $\xi $ along the line (\ref{line}),  for
$\alpha_1=\alpha_2=0$, $\alpha_3=\sqrt{6}$, and $\beta =2\sqrt{6}$. Note
that the apparent singular point at $\xi =\pm\sqrt{2/3}=0.82$ is in fact
just a regular point, as $W$ is the norm of a vector. We indicate the
corresponding values of $x^5$ for the flow. \label{fig:potbc}}
\end{center}
\end{figure}

The critical point (\ref{basepoint}) appears as isolated whenever $|\beta|
\neq \alpha$, where $\alpha>0 $ is the length of the vector $\alpha_r$. If
$|\beta| =\alpha$, then there is a two-dimensional plane where $K=0$. More
precisely, we have to distinguish 3 regions of the parameter space
$\{\alpha_r,\beta\}$.
\begin{enumerate}
  \item If $|\beta| <\alpha $, then the base point is an UV critical point.
  \item If $|\beta|=\alpha $ then there is a plane of critical points.
This plane is parametrized by $\theta$ and $\tau$ with
\begin{eqnarray}
V &=& 1 - \theta^2 - \tau^2 + \frac{2(\alpha_1 \theta - \alpha_2 \tau)}
{\alpha_3 + \beta}\,, \nonumber\\
\sigma &=& - \frac{2(\alpha_2 \theta + \alpha_1 \tau)}{\alpha_3+ \beta}\,.
\end{eqnarray}
In the case $\beta =-\alpha _3$ the plane is $\theta =\tau =0$.  In
  the orthogonal directions, $W$ is increasing, i.e., this plane is of the
  UV type.
  \item If $|\beta |>\alpha $, then $W$ decreases in two directions. The
  critical point is an IR critical point for flow in these directions. In
  this case, $W$ also has a line of zeros. For example, for gauging in the
  direction $\alpha =\alpha_3$ we find the zeros for
\begin{equation}
  V=\frac{\beta -\alpha }{\alpha +\beta }\,,\qquad \sigma =0\,,\qquad \theta
  ^2+\tau ^2=\frac{2\alpha }{\beta +\alpha }\,\qquad \mbox{for }\beta >\alpha
  >0\,.
  \label{zerostoy}
\end{equation}
\end{enumerate}
The last case shows the aspects of the toy model of the universal
hypermultiplet that were not known for vector multiplets only. Let us now
choose $\alpha _1=\alpha _2=0$, which we can safely do due to the SU(2)
invariance. The superpotential can be written as
\begin{equation}
  W=\frac{1}{8\sqrt{6}V}\left( \left\{ (\alpha +\beta )\left[1+\sigma ^2+(V+\xi
  ^2)^2\right] +2(3\alpha -\beta )(V-\xi ^2)\right\} ^2+128\alpha (\alpha -\beta
  )\xi ^2V\right) ^{1/2}\,,
 \label{W34}
\end{equation}
where
\begin{equation}
  \xi ^2\equiv \theta ^2+\tau ^2\,.
 \label{xi2}
\end{equation}
We plot in figure~\ref{fig:conpl34}
contours of constant $W$, where the base point is in the middle. One sees
that $W$ increases in the vertical direction from this point. The central
line in the horizontal direction represents the locus
\begin{eqnarray}
  &&\sigma =0\,,\qquad V+\xi ^2=1\,,\qquad \mbox{with }0<V\leq 1\mbox{ or }
  -1<\xi <1\,, \nonumber\\
  &&\mbox{or }\ V=1-\tanh ^2\chi \,,\qquad \xi =\tanh \chi \,,\qquad \mbox{with }
  -\infty <\chi <\infty \,.
 \label{line}
\end{eqnarray}
In view of (\ref{xi2}), this ``line'' is actually a plane in the full
quaternionic manifold, but as we often use the parametrization
(\ref{line}) we will call it a line. The parametrization by $\chi $,
involving a hyperbolic function, was introduced in~\cite{FGPW}, where
this variable was called $\varphi_1$. On this line, the first two
components of the Killing vector are zero, and the beta functions of the
two field combinations vanish:
\begin{equation}
    \beta^\sigma = \beta^{V+ \xi^2} = 0\,.
    \label{eq:betas}
\end{equation}
Here, the Killing vector has as the only nonzero components $K^\theta
=(\alpha -\beta )\tau /2$ and $K^\tau =-(\alpha -\beta )\theta  /2$.
These are indeed never vanishing for $\beta \neq \alpha $ except at the
base point (\ref{basepoint}). On this line, the superpotential reduces to
\begin{equation}
  W_{\rm line}=\frac{|\alpha (2-\xi^2)-\beta
\xi^2|}{2\sqrt{6}(1-\xi^2)}\,.
 \label{Wline}
\end{equation}
This means that there is a zero for
\begin{equation}
 \xi ^2=\frac{2\alpha }{\alpha +\beta }\,,
 \label{zeroxi}
\end{equation}
which is in the domain of definition if $\beta >\alpha>0$. We plot the
superpotential on this line for such a case in figure~\ref{fig:potbc},
exhibiting the zeros, which are in a circle in the $(\theta ,\tau )$
plane.

We will now consider a flow on that line. The critical point at $\xi =0$
is used for the asymptotic values of the zeros at $x=+\infty $ and the
zero of the superpotential is placed at $x^5 = 0$. The BPS equation
(\ref{consistgW}) leads on both sides of the zero to (we put here $g=1$)
\begin{equation}
  A'=\frac{\alpha (2-\xi ^2)-\beta \xi ^2}{2\sqrt{6}(1-\xi ^2)}\,.
 \label{exAprime}
\end{equation}
Note that the sign flip has disappeared. For the other BPS equation
(\ref{qprimeall}), we use the inverse metric to obtain\footnote{
Alternatively, one can use that the metric reduced to the line is $\rmd
s^2 = \frac{2}{V^2} \left(\rmd \tau^2 + \rmd \theta^2\right)$, to obtain,
e.g., $\tau '=-\ft{1}{2}\sqrt{\ft{3}{2}} (\beta-\alpha)\tau$, also
leading to (\ref{eq:Vprime}).}
  to obtain
\begin{equation}
    V^\prime = 3\partial^VW=\sqrt{\ft{3}{2}} (\beta-\alpha)\xi^2
    =\sqrt{\ft{3}{2}} (\beta-\alpha)(1-V)\,,
    \label{eq:Vprime}
\end{equation}
where again the sign flipping has disappeared.

Choosing the integration constant such that the zero is $x^5 = 0$ and the
warp factor reaches the value $a(0)=1$, the solution is given by
\begin{equation}
   \xi^2(x^5)=\frac{2\alpha }{\alpha+\beta} \; e^{-\sqrt{3/2} \,
    (\beta-\alpha)\,x^5}\,,\qquad
    A(x^5)=-\frac{\alpha \,x^5}{{\sqrt{6}}}
   + \frac{1}{6}\log \frac{\alpha+\beta-2\alpha e^{- \sqrt{3/2}
           \, ( \beta-\alpha) \,x^5}}{\beta -\alpha }\,.
    \label{eq:Ainx}
\end{equation}
This flow has a singular point at $x^5 = -\sqrt{\frac{2}{3}}
\frac{1}{\beta-\alpha} \log \frac{\alpha+\beta}{2\alpha}$, where the
border of the quaternionic manifold is indeed reached as $\xi^2=1$. Here
$A \to -\infty$. At the other end, one reaches the fixed AdS critical
point for $x^5 \to + \infty$, where the behaviour of the warp factor is
$A \to -\frac{\alpha}{\sqrt{6}} x^5 = - W_{\rm cr} x^5$, which is the
asymptotic of an IR AdS fixed point.

\begin{figure}\unitlength=1mm
\begin{center}
\begin{picture}(120,50)(0,0)
\put(20,0){\leavevmode \epsfxsize=7cm
 \epsfbox{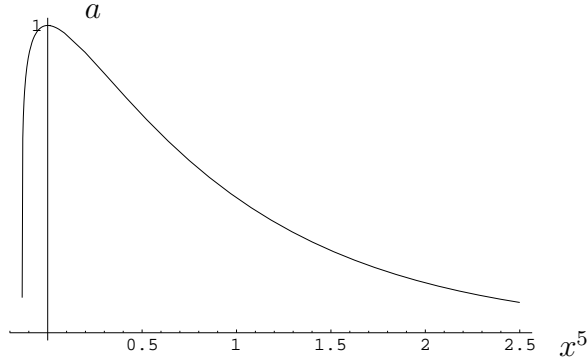}}
 \put(93,0){$x^5 $}
 \put(30,45){$a$}
\end{picture}
\caption{\it Graph of the warp factor $a(x^5)=e^A$ for the values
$\alpha=\sqrt{6}$ and $\beta =2\sqrt{6}$. \label{fig:warp}}
\end{center}
\end{figure}

We display in figure~\ref{fig:warp} the exponential of the warp factor for
such a solution, showing that it is perfectly well behaved and continuous
at the point $x^5 = 0$ where it reaches its maximum equal to~1.

As $W$ has no other critical points, one can have zeros only in that
case. In~\cite{Behrndt:2000ph}, where the parametrization
of~\cite{Britto-Pacumio:1999sn} was used, some misleading results were
obtained. We discuss this in appendix~\ref{app:BC}.

These features regarding the case with one hypermultiplet and no vector
multiplets can be generalized to arbitrary quaternionic K{\"a}hler
homogeneous spaces $\mathrm{G}/(\mathrm{SU}(2)\times \mathrm{K})$. There
can only be one connected region with critical points, and the
eigenvalues of the Hessian are always spread evenly from~$3/2$ as in
(\ref{EigenddW}). To get IR directions in the critical point the gauging
has to be ``mainly'' in the direction of K, while some gauging in the
direction SU(2) is necessary for a nonzero $W$ at the critical point. All
this will be shown in~\cite{Alekseevsky:2001if}.

\subsection{The full model and the FGPW flow}

We now consider the full model, with  a vector multiplet and a
hypermultiplet. Taking into account the graviphoton, we have two vector
fields, and therefore we can gauge two Abelian isometries $U(1)\times
U(1)$. As explained before, we should find the FGPW flow between an UV
and an IR fixed point by a specific $\mathrm{U}(1)\times \mathrm{U}(1)$
gauging.

We will start by considering the requirements imposed by the presence of
\textit{a first critical point}, for which we will choose the base point
\begin{equation}
\mbox{c.p. }1:\ q=\{1,0,0,0\}\,, \qquad \rho = 1\,.
 \label{cr1}
\end{equation}
We must first solve the condition $K_{\rm c.p.1} = (h^0 K_{0}+ h^1
K_{1})_{\rm c.p.1} = 0$.  As in the toy model and as a general result,
this implies that $K_{\rm c.p.1}$ generates a U(1) inside the isotropy
group of the scalar manifold.  In our specific case, this results in the
absence of contributions from the noncompact generators in the
combination $K_0+\sqrt{2}K_1$. Furthermore, we can use the SU(2)
invariance to choose just one direction in SU(2). We again take the
generator $T_3$.

Since we now also have a vector multiplet scalar, the critical points of
the superpotential are defined by both attractor equations
(\ref{fixedpoints}), and one must also satisfy the requirement for the
prepotentials $h_I P^r= P^r_I$. Due to the self-consistency of this
equation (multiplying it by $h^I$), this gives only one triplet of
requirements:
\begin{equation}
  h_0P_1^r=h_1P_0^r \qquad \mbox{at the critical point.}
 \label{requirhP}
\end{equation}
Only the generators of the SU(2) part of the isotropy group contribute to
the prepotentials, and the 3 generators have 3 independent prepotentials.
Therefore, this condition does not lead to any constraint on the
noncompact generators. However, it does give conditions on the generators
in the SU(2) part of the isotropy group that imply the absence of $T_1$
and $T_2$ from $K_0$ as well as $K_1$. Furthermore, it fixes the relative
weight of $T_3$ in both generators. As a result of the attractor
equations at c.p. 1, we can parametrize the generators as
\begin{eqnarray}
 K_0 & = & \sqrt{2}\alpha\left( \frac12\, T_3+\frac1{\sqrt{3}}\gamma\, T_8
  +T_{\rm nc}\right)\,, \nonumber\\
 K_1 & = & \alpha\, \left( T_3+\frac{1}{\sqrt{3}}\beta\, T_8-T_{\rm nc}\right) \,,
 \label{K01}
\end{eqnarray}
where $\alpha >0$ and $T_{\rm nc}$ is a linear combination of the
noncompact Killing vectors.

In order to obtain two independent U(1)'s we still must impose the
requirement that the two generators commute. It can be easily seen that
this is equivalent to requiring a vanishing commutator of $T_{\rm nc}$
with the combination $\frac{3}{2}T_3+\frac{\beta +\gamma} {\sqrt{3}}T_8$.
Thus for a general quaternionic manifold the generators whose roots lie in
the direction defined by this generator are the ones that can survive, if
there are any. Therefore noncompact generators are possible only if
$\beta +\gamma= \pm\frac{3}{2}$. We formulated the analysis such that it
is suitable for an easy generalisation to situations with one vector
multiplet and an arbitrary homogeneous quaternionic manifold.

As the prepotentials depend only on the SU(2) part, we find that the
value of $W$ at the critical point is proportional to $\alpha$:
\begin{equation}
  W_{\rm c.p.1}=\ft12\alpha \,.
 \label{Wcp1}
\end{equation}

We now turn to analyse the (IR/UV) properties of the critical point,
computing the eigenvalues of the matrix (\ref{calUres}). In the matrices
in (\ref{defcalJL}), ${\cal J}$ depends only on the gauging in the SU(2)
direction, i.e., it is proportional to $\alpha $, while ${\cal L}$ is a
function of the U(1) gauging, i.e., it is proportional to $\beta +\gamma$.
This implies that in the generic case we find
\begin{equation}
{\cal U}_{\rm c.p.1}= \diag\left\{\ft{3}{2}+\beta +\gamma,\,
  \ft{3}{2}+\beta +\gamma,\,
  \ft{3}{2}-\beta -\gamma,\,
  \ft{3}{2}-\beta -\gamma,\,2\right\}.
 \label{Ucp1}
\end{equation}
Thus, noncompact generators can enter when this matrix has zero
eigenvalues. From the form of (\ref{calUres}), one can see that the
noncompact symmetries can contribute only to the off--diagonal elements
that are in the direction of the zero modes of the pure hypermultiplet
part of ${\cal U}$. They therefore modify the part $(0,0,2)$ of
(\ref{Ucp1}). The result is that the eigenvalues are then
\begin{equation}
  \eig {\cal U}_{\rm c.p.1}=\left\{ 3,3,0,1+\sqrt{1+6a^2},1-\sqrt{1+6a^2}\right\}\,,
 \label{eigUcp1}
\end{equation}
where $a$ is the weight with which the noncompact generators appear (e.g.,
$T_{\rm nc}=aT_4$).

\medskip

We should remark that the critical point c.p. 1 does not in general
preserve the full $\mathrm{U}(1) \times \mathrm{U}(1)$ gauged symmetry.
Indeed, the gauge invariance could be spoiled by the presence of a mass
term for the gauge vector coming from the kinetic part of the hyperscalars
(\ref{massterm}). Since the mass is related to the norm of the Killing
vectors of the gauged isometries, we see that the invariance is broken
anytime $K_{I}^{X} K_{IX} \neq 0$. This happens whenever we turn on the
noncompact generators. In our case, $T_{\rm n.c.} \neq 0$ implies that at
the c.p. 1 the $\mathrm{U}(1) \times \mathrm{U}(1)$ isometry is broken to
the U(1) generated by $K_0 + \sqrt{2} K_1$.

As our present interest is specifically aimed at reproducing the FGPW
flow, from now on we restrict consideration to $T_{\rm nc}=0$. This
allows the existence of a single point where the full $\mathrm{U}(1)
\times \mathrm{U}(1)$ gauge symmetry is preserved.

\medskip

We now go back to analyzing the \textit{BPS flows} that can originate
from c.p. 1. Remember that with mixed vector and hypermultiplets they have
to satisfy the requirement
\begin{equation}
  \partial _\rho Q^r=0\,.
 \label{drhoQ}
\end{equation}
This is not satisfied for a generic point in the manifold. However, it is
satisfied on the line (\ref{line}) and thus we further consider a flow
along it. In order for the flow to be consistent with the restriction,
the flow equation (\ref{qprimeall}) should not drive the fields off the
line and this translates into requiring that
$\beta^{\sigma}=\beta^{V+\xi^2} = 0$. These conditions are indeed
satisfied. Along the line, the quantities $K$, $W$, and $Q$ simplify
to\footnote{The $\pm$ in the expression for $Q^r$ is dependent on whether
the expression in $W$ of which we have to take the modulus is positive or
negative.}
\begin{eqnarray}
  \left.K^X\right|_{\rm line}&=&\frac{\alpha}{\sqrt{6}\rho ^2}\left[1  -\beta
   +\rho ^6\left(\frac{1}{2} -\gamma\right)\right]
  \,(0,0,\tau ,-\theta )\,,
 \nonumber\\
 \left. W \right|_{\rm line}&=&\frac{\alpha }{6\rho ^2(1-\xi ^2)}
 \left|(2+\rho ^6)
 (1 -\ft12\xi ^2 )-(\beta +\gamma \rho ^6)\xi ^2
 \right| , \label{KWQline}\\
 \left.Q^r\right|_{\rm line}&=&\pm\left(2\theta \sqrt{1-\xi ^2},
 2\tau \sqrt{1-\xi ^2}, -1+2\xi ^2 \right) .
\nonumber
\end{eqnarray}
\medskip

We now fix the end point of the flow at \textit{a second critical point}
and we have again two requirements: the vanishing of the Killing vector,
and the requirement (\ref{requirhP}). These now fix the values of $\rho$
and $\xi^2$ for the critical point.

The value of $\rho$ for the second critical point follows directly from
the first of (\ref{KWQline}), while we use (\ref{requirhP}) to fix also
$\xi^{2}$. Along the line, the prepotentials are given by
\begin{equation}
 \left. \pmatrix{P_0^r\cr P_1^r}\right|_{\rm line}= \frac{\alpha\,Q^r}{4(1-\xi ^2)}
  \pmatrix{\frac{1}{\sqrt{2}}\left( 2-\xi ^2-2\gamma \xi
  ^2\right)
  \cr  2-\xi ^2-\beta  \xi ^2 }\,.
 \label{PIline}
\end{equation}
Note that they now involve a $\xi $-dependent mixing of the parameters
that appear in (\ref{K01}), but are all proportional to the same $Q^r$.
This comes about because, as we mentioned before, the relevant part is the
SU(2) part of the isotropy group. This isotropy group rotates within the
full SU(2,1) while we move over the line. With only $T_3$ and $T_8$ used
in our gauging, the effective SU(2) still has its first two components
zero. That is why all entries in (\ref{PIline}) are proportional to the
same matrix $Q^r$. The third component of the SU(2) part of the isotropy
group is a linear combination of $T_3$ and $T_8$, leading to the mixture
in (\ref{PIline}). Using $h_1/h_0=\sqrt{2}\rho ^6$, we find for the second
critical point\footnote{As mentioned above, we always have a full circle
of critical points for these values of $\theta_{\rm cr}$ and $\tau_{\rm
cr}$. }
\begin{eqnarray}
\mbox{c.p. }2: && q=\{V_{\rm cr},0,\theta_{\rm cr} ,\tau _{\rm cr}\}\,,
\qquad \rho =\rho _{\rm cr}\,,
\qquad \rho_{\rm cr}^6 = \frac{2(\beta -1) }{1 -2\gamma }\,,  \label{cr2abc}\\
   &&
   \xi ^2_{\rm cr}=
   \theta_{\rm cr}^2+\tau ^2_{\rm cr}
   =\frac{2 (1-\rho_{\rm cr}^6)} {3\beta -1-2\rho_{\rm cr} ^6 }
   =\frac{6  -4 (\beta +\gamma )}
  {3-  \beta  + 2\gamma - 6\beta \gamma }
    \,, \qquad
    V_{\rm cr}
    = \frac{3(\beta- 1 )}{3\beta -1-2\rho_{\rm cr}^6 }
    \nonumber\,.
\end{eqnarray}
We find for the value of $W$ at this second critical point
\begin{equation}
  W_{\rm c.p.2}=\frac{\alpha (\beta -2\gamma )}
  {3(1 -2\gamma )\rho_{\rm cr} ^2}
  =\frac{\alpha (2+\rho_{\rm cr} ^6)}{6\rho_{\rm cr} ^2}\,,
 \label{Wcp2}
\end{equation}
such that
\begin{equation}
 \frac{W^3_{\rm c.p.1}}{W^3_{\rm c.p.2}}=\frac{27\rho_{\rm cr}
  ^6}{(2+\rho_{\rm cr} ^6)^3}= \frac{27(1-\beta )(1-2\gamma )^2}{4(2\gamma -\beta )^3}\,.
 \label{W3ratio}
\end{equation}

The condition that the critical point is in the domain can be written as
\begin{equation}
  (\rho_{\rm cr} -1)(1 -\beta )>0\,.
 \label{condcp2}
\end{equation}
Notice that this is exactly the condition that the third and fourth
entries of (\ref{Ucp1}) ($\theta $ and $\tau $ directions) are positive.
These are the ones that are relevant when one computes the eigenvalues of
the matrix (\ref{calUres}) restricted to the line, and \textit{excludes
the possibility that the first critical point has IR directions along the
flow}.

To summarize, we have presented above a two-parameter model, where $\beta
$ and $\gamma $ are free parameters and $\alpha $ defines an overall
normalization related to the AdS radius.

\medskip

We will now give more details of the models for specific values of the
parameters, leading also to the identification of the FGPW potential as a
part of our two-parameter model. We will further restrict consideration to
the branch $\rho >1$ and thus $\beta <1$. For definiteness we now choose
the value of $\rho $ at the second critical point as in~\cite{FGPW}:
\begin{equation}
  \rho ^6_{\rm cr}=2\,\qquad \Rightarrow \qquad
  \gamma = 1 - \frac{\beta}{2}\,.
 \label{valgamma}
\end{equation}
With this choice, the second critical point is thus at
\begin{eqnarray}
\mbox{c.p. }2: && q=\{V_{\rm cr},0,\theta_{\rm cr} ,\tau _{\rm cr}\}\,,
\qquad \rho_{\rm cr} =  2^{1/6}\,, \nonumber\\
   && V_{\rm cr} = \frac{3(1-\beta)}{5-3\beta}\,, \qquad
   \xi ^2_{\rm cr}=\theta_{\rm cr}^2+\tau ^2_{\rm cr}= \frac{2 }{5 -3\beta
   }\,.
    \label{cr2}
\end{eqnarray}
We are therefore left with a one-parameter family of models with two
critical points where $\beta$ fixes the ratios of the gaugings. A second
critical point appears only if $\beta <1 $.

We can also compute the value of the cosmological constant $W^2$ at the
critical points:
\begin{equation}
W^2_{\rm c.p.1} = \dfrac{\alpha^2}{4}\,,  \qquad W^2_{\rm c.p.2} =
\dfrac{2^{4/3}}{9} \alpha^2\,, \qquad \frac{W^3_{\rm c.p.1}}{W^3_{\rm
c.p.2}}=\frac{27}{32}\,;
 \label{Wcr12}
\end{equation}
the relation that is found also in~\cite{FGPW}, and which we generalize
here for arbitrary $\beta$.

\medskip

Now that we have two critical points we still have to discuss their
nature. As we have already excluded the possibility that the first
critical point has IR directions along the flow, we are interested in
whether these models have interesting applications for renormalization
group flows. We are indeed interested in having an UV and an IR critical
point such that we can reproduce the FGPW flow.

To understand the nature of such points we have to look again at the
Hessian matrices of the superpotential and at their eigenvalues. At the
first critical point (\ref{cr1}), the $\cU$ matrix (\ref{Ucp1}) reduces to
\begin{equation}
 {\cal U}_{\rm c.p.1}= \diag\left(\ft{5}{2}+\ft12{\beta},\
    \ft{5}{2}+\ft12{\beta},\
\ft{1}{2}-\ft12{\beta},\ \ft{1}{2}-\ft12{\beta},\ 2\right).
    \label{eq:eig1}
\end{equation}
The value of the vector scalar sector is still the characteristic one of
the very special vector scalar manifold \cite{KL,BC}. The values in the
hypermultiplet sector follow instead the pattern outlined by the formula
(\ref{calUres}). As mentioned already, the third and fourth entries are
positive if we demand the existence of the second critical point ($\beta
<1 $). If, in addition, one satisfies the more stringent constraint
$\beta <-5$, then there are two IR directions, which, however, are not
along the flow.

At the other critical point the eigenvalues are given by
\begin{equation}
   \eig {\cal U}_{\rm c.p.2}=  \left\{0,3,3,
1+\ft{1}{2}\sqrt{19 - 9 \beta}, 1-\ft{1}{2}\sqrt{19 - 9 \beta}\right\},
    \label{eq:eig2}
\end{equation}
no matter which point is chosen on the critical circle. With the limit
$\beta <1 $ that we already obtained, this implies that there is always
one IR direction. If the first critical point has an IR direction $\beta
<-5 $, then both critical points have an IR direction. This is thus the
first example of a model with two IR fixed points. However, the flow
along the line that we consider does not connect in these directions. It
is difficult to indicate a flow along another line that would connect the
two, or to exclude this possibility.

Note that (\ref{eq:eig2}) is of the form (\ref{eigUcp1}), with
$a^2=(15-9\beta )/24$. This can be understood as follows. As we stressed
before, any point in the manifold is equivalent. Thus what we find at the
second critical point should fit in the pattern that we discussed for the
first critical point. However, the generators $T_3$ and $T_8$ that appear
in (\ref{K01}) are not in the isotropy group of any other point of the
manifold. Thus, in this second critical point, the generator $K_I$ with
$T_{\rm nc}=0$ should be interpreted as having a part $T'_3$ [giving rise
to (\ref{PIline})], a part $T'_8$, and a part $T'_{\rm nc}\neq 0$.
Therefore this second critical point should fall in the general analysis
for a critical point with gaugings outside the isotropy group, i.e., the
eigenvalues should be of the form (\ref{eigUcp1}) where $a$ measures the
amount in which the generators $T_3$ and $T_8$ contribute to $T'_{\rm
nc}$. This principle could be used for an alternative, group-theoretical,
analysis of the possibilities for the second critical point.

The contribution of $a^2=(15-9\beta )/4$ to the eigenvalues arises in
${\cal U}$ by the mixing of the scalars of the vector multiplet and
hypermultiplets.

This means that when vector multiplets are added to hypermultiplets, the
appearance of IR directions can be due to two different mechanisms. One is
the presence of the hypermultiplets, which allows values as in
(\ref{eigUcp1}), which would lead here to $(3,3,1,1,0)$. The other is due
to the possible breaking of the gauge symmetry, which occurs if their
generators are outside the isotropy group of the critical point, such
that $K^Z_I$ in the off-diagonal elements of (\ref{calUres}) is
nonvanishing. This shifts the $(1,1)$ eigenvalues to the values in
(\ref{eq:eig2}).

\bigskip

The resulting form of the superpotential is, using the parametrization in
(\ref{line}),
\begin{equation}
    W = \frac{\alpha }{4\rho^2}\left[1
    +\frac{1}{3}\beta +\frac{\rho^6}{6}(5-\beta)- \frac{(1-\beta)}{6}
    (\rho^6 - 2) \cosh(2\chi)
    \right].
\end{equation}
It is striking that for the choice $\alpha = 3$ and $\beta = -1$ one can
retrieve the superpotential\footnote{In~\cite{FGPW} $W$ was chosen to be
the definition $W = - |W|$ rather than the one given in (\ref{Wmydef}),
and thus differs by a sign from ours. To compare our flow equations with
those of~\cite{FGPW}, we have to take our coupling constant to be $2/3$
rather than 2, as in~\cite{FGPW}.} presented in~\cite{FGPW}
\begin{equation} \label{potFGPW} W =- \frac{1}{4\rho^2}\left[
\left(\rho^6 - 2\right) \cosh(2 \chi) - \left(2
    +3 \rho^6\right)\right],
\end{equation} and therefore also the flow.
For this value of $\beta$ we have indeed one IR and one UV critical point:
 \begin{eqnarray}
\textrm{UV}: && \mbox{c.p. }1 :\ \rho = 1\,, \quad\hspace{5mm} \xi =
0\,,\quad
\hspace{3mm}(\chi =0)\,,  \nonumber\\
\textrm{IR}: && \mbox{c.p. }2 :\ \rho = 2^{1/6}\,, \quad \xi =
\pm\ft12\,,\quad (\chi =\pm\ft12\log 3)\,,\label{crFGPW}
\end{eqnarray}
and the value of $W$ at these critical points is $W_{\rm c.p.1} =
\frac{3}{2}$ for the UV and $W_{\rm c.p.2} = 2^{2/3}$ for the IR, both
representing AdS vacua.

\subsection{FGPW flow in $\cN=2$ theory}

Let us now give a closer inspection to the FGPW model embedded in $\cN=2$
theory. In terms of the scalar manifold isometries, the FGPW flow can be
retrieved by gauging those generated by
\begin{equation}
K_{(0)} \equiv \frac{3}{\sqrt{2}} \,
\left(k_1+k_6\right)=\frac{3}{\sqrt{2}} \, \left(T_3 + \sqrt{3} T_8\right)
\end{equation}
and by
\begin{equation}
K_{(1)} \equiv - 3 \, k_4=\sqrt{3}\left(\sqrt{3} T_3 - T_8\right).
\end{equation}

The corresponding superpotential (in terms of all the coordinates)
depends on $\theta $ and $\tau $ via the combination $\xi $ in (\ref{xi2})
and is
\begin{eqnarray} W &=& - \frac{1}{4\,\rho^2\,V} \left\{16\,
{\left( V + \xi ^2 \right) }^2 +  2\,\sigma^2\,\rho^{12}\left[ 1 + \left(
V +
\xi ^2 \right)^2\right] +\right. \nonumber \\
&+& \rho^{12}\,\left[ \sigma^4 + V^4 +  4\,V^3\,\xi ^2 + {\left( 1 + \xi
^4 \right)
  }^2 +2\,V^2\,\left( 1 + 3\xi ^4 \right)\right] + \label{W}\\
&+&\left.  8\,\rho^6\, \left( V - \xi ^2 \right) \, \left[ 1 + \sigma^2 +
\left(V + \xi ^2\right)^2 \right] + 4\,V\,\rho^{12}\left[ \xi ^6
 -3\xi ^2
\right]  \right\}^{1/2}\,.\nonumber
\end{eqnarray}
It is therefore more convenient to study its stationary points in the
variables $\{V,\sigma,\xi,\rho\}$.

A general analysis of the (\ref{W}) gives indeed only the two expected
critical points at
\begin{equation}
\mbox{c.p. }1 :\ \sigma = \xi^2 = 0\,,\qquad V = 1\,,\qquad \rho \equiv
e^{\phi_3 / \sqrt{6}} = 1 \label{UV}
\end{equation}
 and at
\begin{equation} \label{IR}
\mbox{c.p. }2 :\ \sigma = 0\,,\qquad  V = \ft{3}{4}\,,\qquad  \rho =
2^{1/6}\,,\qquad \xi^2 = \ft{1}{4}\,.
\end{equation}

The first is the expected isolated UV point. As explained above, we can
identify it as an UV fixed point by considering the leading contributions
to the $\beta$ function of the couplings $\phi =\{V, \sigma, \theta,
\tau$;$\rho \}$ that are encoded in the eigenvalues of the matrix
(\ref{defcU}),
\begin{equation}
{\cU}_{\rm c.p.1}  = \left(
\begin{array}{ccccc} 2 &&&&\\ & 2 &&& \\ &&
    1 &&\\&&&
    1& \\ &&&& 2 \end{array} \right),
\end{equation}
which are all positive.

The second  critical point (\ref{IR}) actually represents  a whole circle
of saddle points. We show here the explicit form of the Hessian matrix at
the point $\theta = \frac{1}{2}$ and $\tau = 0$:
\begin{equation}
\cU_{\rm c.p.2}  = \left( \begin{array}{ccccc} \frac{9}{4}  && \frac{3}{4}
&&
\frac{2^{1/6}}{3}\\
&  3 && \frac{3}{4}&\\
\frac{9}{4} && \frac{3}{4}& & - 2^{1/6}\\
&&&0 &\\
\frac{9}{2^{7/6}} && -\frac{9}{2^{7/6}}& &2
\end{array} \right), \label{Hessianex}
\end{equation}
whose eigenvalues are $0$,$3$,$3$,$(1+\sqrt{7})$ and $(1-\sqrt{7})$,
which is (\ref{eigUcp1}) with $a=1$.

As foreseen above, here the mechanism that determines the appearance of an
infrared direction is different from the one shown for the
hypermultiplets alone. In this case indeed the negative eigenvalue comes
from the mixed partial derivatives with respect to the hyperscalars $V$,
$\theta$ and the vector multiplet scalar $\rho$. The null eigenvalue is
related to the massless Goldstone scalar that is eaten by the vector
combination which becomes massive at this point.

We also want to point out here that the presence of a whole line of
critical points should be connected to the fact that in the dual CFT one
expects to have a line of exactly marginal perturbations \cite{LS} to the
theory at such an IR point.

\bigskip

After identification of the correct UV and IR end points, we now turn to
the second important guideline for the identification of the FGPW
discussed in section~\ref{ss:scManIsom}, which was related to the mass
terms for the gauged vector fields, and thus to the norm of the gauged
Killing vectors. In the FGPW example, at the UV fixed point both the
graviphoton and the gauge vector are massless, whereas at the IR point
only the graviphoton is still  massless and is the gauge vector of the
residual U$(1)_R$ symmetry.

In order to translate these facts into our present language, we observe
that all along the flow where $\sigma=0$ and $V+\xi^2=1$, the two Killing
vectors $K_{(0)}$ and $K_{(1)}$ are proportional to one another and are
equal to
\begin{equation} \label{rela} K_{(1)} = - \sqrt{2} \,
K_{(0)} = 3 \, \left(\begin{array}{c} 0 \\ 0 \\ \tau \\
-\theta
  \end{array} \right),
\end{equation}
and this translates the statement that along the flow the combination $V +
\theta^2 + \tau^2$ and the $\sigma$ field should remain constant. It is
straightforward to see that $\delta_{K_{(0)}} (V + \theta^2 + \tau^2) =
\delta_{K_{(0)}} \sigma = 0$ and the same for $K_{(1)}$.

Equation (\ref{rela}) then allows us to identify the graviphoton with the
gauge vector of the U$(1)_R \subset \textrm{U}(1) \times \textrm{U}(1)
\subset \textrm{SU}(2) \times \textrm{U}(1)$ symmetry generated by
\begin{equation} K_R \equiv \sqrt{2} \, K_{(0)} + K_{(1)}\,\ \left(
= 0 \hbox{ along the flow}\right).\label{KR}
\end{equation}

Let us then analyse the relevant supersymmetry transformations at the IR
and UV fixed points in order to identify which mixture of the vector
fields ${\cal A}_\mu^I$ gives rise to the graviphoton and to the extra
vector field.

\bigskip

We find that upon defining
\begin{eqnarray}
\label{cA}
\cA_\mu &\equiv&\frac{2^{1/3}}{\sqrt{6}}\left(
\frac{A_\mu^0}{\sqrt{2}} + 2 \, A_\mu^1 \right),\\
\label{cB}
\cB_\mu &\equiv& \frac{2^{1/3}}{\sqrt{6}}\left(
\sqrt{2} A_\mu^1 - A_\mu^0 \right),
\end{eqnarray}
the SUSY transformations at the IR point reduce to
(at leading order in the Fermi fields)
 \begin{eqnarray}
\label{susy1} &&\delta_{\epsilon} \, \psi_{\mu \,i} = D_\mu \epsilon_i +
\frac{\rmi}{4\sqrt{6}} \left(\gamma_{\mu\nu\rho} \epsilon_i - 4 g_{\mu\nu}
\gamma_\rho \epsilon_i\right) \,
\cF^{\nu\rho} + \cdots, \\
&&\delta_{\epsilon} \cA _\mu = \rmi \frac{\sqrt{6}}{4}\bar{\psi}^i_\mu
\epsilon_i\,, \nonumber\\
\label{susy3} &&\delta_{\epsilon} \cB_\mu = -\frac{1}{2} \,
\bar{\epsilon}^i \gamma_\mu \lambda_i\,,
 \end{eqnarray}
where we have also defined $\cF = d \cA$.
Therefore, by these equations one identifies the graviphoton field with
the $\cA_\mu$ combination, and the vector at the head of the massive
vector multiplet with  $\cB_\mu$.

\textit{At this IR point}, for the mechanism we showed, the true massive
vector $\cB_\mu$ is given by an appropriate sum of (\ref{cB})  and $D_\mu
q$, where $q$ is the right combination of the hyperscalars that acts as a
Goldstone boson. Therefore the full supersymmetry transformation rule for
$\cB_\mu$ will also contain a term of the type $D_\mu\left(
\delta_{\epsilon} q \right).$

If we associate with the graviphoton its Killing vector proportional to
$K_R$, we find that the U$(1)_R$ symmetry gauged by the graviphoton is
generated by
\begin{equation}
K_{\cA} = \frac{2^{2/3}}{\sqrt{6}} \, K_R,
\end{equation}
while the generator of the (broken) U$(1)_{\cal B}$ isometry associated
with the massive vector is given by
\begin{equation}
K_{\cB} = \frac{2^{2/3}}{\sqrt{6}} \,
\left(\frac{1}{\sqrt{2}} K_{(1)} - 2
  K_{(0)} \right).
\end{equation}

\bigskip

\textit{At the UV fixed point} one can again rewrite the supersymmetry
rules as in equations (\ref{susy1}) and (\ref{susy3}), provided that now
one makes the identifications
\begin{eqnarray} \cA_\mu &=& \frac{1}{\sqrt{3}} \left(  A_\mu^0 +
\sqrt{2} A_\mu^1
  \right), \\
\cB_\mu &=& \frac{1}{\sqrt{3}}\left(  A_\mu^1 -
\sqrt{2}A_\mu^0 \right).
\end{eqnarray}
Again, the graviphoton field is identified with the $\cA_\mu$
combination, and the gauge vector with $\cB_\mu$.

If  we relate to these vectors the Killing generators of the U(1)
isometries that they gauge [we remark that at this point they are both
massless and that both are gauge vectors of U(1) isometries], one can see
that the graviphoton gauges the U(1) generated by
\begin{equation} K_{\cA} = \frac{1}{\sqrt{3}} \, \left(K_{(0)} + \sqrt{2}
K_{(1)}\right), \end{equation} whereas the massless vector gauges that
generated by
\begin{equation} K_{\cB} = \frac{1}{\sqrt{3}} \, \left(K_{(1)} - \sqrt{2}
K_{(0)}\right). \end{equation} This means that along the flow the $R$
symmetry gauged by the graviphoton has rotated.

The interpretation of this fact in terms of the dual CFT \cite{FGPW} is
the following. When one adds a mass term to the CFT at the UV point, the
$R$ current connected to the graviphoton becomes anomalous (i.e., the
original graviphoton acquires a mass along the flow) whereas the
nonanomalous one is the combination that keeps the original graviphoton
and the other gauge vector massless. As stated in~\cite{DW}, this latter
couples to the Konishi current, and therefore the one that couples to the
anomaly-free U$(1)_R$ current must be a combination of the original
graviphoton and this latter.

In detail, at the UV fixed point, both $K_{\cA}^{UV}$ and $K_{\cB}^{UV}$
are equal to 0, but, as we move away from it, $\theta$ and (or) $\tau$
will vary and then they will no longer be 0. For the sake of simplicity,
choosing the $\theta = \frac{1}{2}$ and $\tau = 0$ point between the IR
points (\ref{IR}), we can also keep $\tau = 0$ for all the flow (indeed,
for such conditions $\beta^\tau = 0$ also, no matter what the $\theta$,
$V$, and $\rho$ values) and parametrize the other variables as $\theta =
\xi$ and $V = 1-\xi ^2$. Then the Killing vector of the broken
U$(1)_{{\cal B}}$ isometry will be parametrized by
\begin{equation}
K_{\cB}= - 3 \sqrt{3}\, 2^{-1/3} \, \left(\begin{array}{c} 0 \\ 0 \\ 0\\
\xi  \end{array}\right),
\end{equation}
and this will therefore give rise to a mass term proportional to
$\frac{1}{2} m^2_{\cB} = \frac{1}{2} \, g_{XY} \, K_{\cB}^X K_{\cB}^Y$
for the $\cB_{\mu}$ vector field (with kinetic term $-\frac{1}{4}
F_{\cB}^2$):
\begin{equation}
\label{mass} m^2_{\cB} =27 \times  2^{1/3}\, g^2\, \frac{\xi ^2 }{(1-\xi
^2)^2} \,.
\end{equation}
This is precisely 0 at the UV fixed point and flows to $6 \times 2^{4/3}
g^2$ at the IR. This means also that at this point its conformal
dimension is given by
\begin{equation}
    E_{0} = 2+ \sqrt{1+\dfrac{m^2}{W^2_{\rm c.p.2}}} = 2 + \sqrt{7},
\end{equation}
which was the expected one for this massive vector~\cite{FGPW}.

\medskip

An interesting example regarding flows in the theory with two
hypermultiplets spanning the
$\frac{\mathrm{G}_{2,2}}{\mathrm{SU}(2)\times \mathrm{SU}(2)}$ manifold
has recently been investigated in~\cite{Bianchi}, where, using $\cN = 8$
supergravity, it has been shown that the effective mass term of the
vector fields reduces to the second derivative of the warp factor.
Although equation (\ref{mass}) does not comply with this request, we
still have to take into account the contribution coming from the
noncanonical normalization of the kinetic term. In fact, we have seen in
the previous section that the fields that have the interpretation of
graviphoton and (gauge) vector, rotate along the flow. Thus, one can still
expect the equation $m^2 \sim A^{''}$ to arise upon performing some
nontrivial field redefinitions.

Presently, the vector field kinetic term has in front a function of the
vector modulus $\rho$. This signals the mixing between the two vectors to
be resolved. One hopes that a suitable rescaling of the vector fields
yielding the standard normalization of the kinetic terms will also do the
job of disentangling this mixing and of giving the correct relation
between the effective mass function and the warp factor.

\section{Discussion and Outlook}\label{ss:conclRem}

Generic gauged  supergravities were not considered relevant from the
perspective of a string theory in a flat background until very recently.
Indeed, due to the discovery of the AdS/CFT correspondence the
AdS${}_5\times \mathrm{X}^5$ background of string theory and the gauged
supergravity in 5D came into the spotlight both in the maximally
supersymmetric ($X^5 = S^5$) and in the lower supersymmetric ($X^5 =
T^{11}$) cases~\cite{T11papers}.

The latter will be useful in trying to better identify the supergravity
details that allow one to select precisely the dual conformal field theory
operators. This will continue the analysis of the previous section along
the lines of \cite{Bianchi}.

\

This paper has uncovered the properties of a general class of 5D $\cN=2$
gauged supergravities, which  have a rich structure of vacua and
interpolating flows.

As a first general result, we have specified  the set of conditions under
which supergravity models coupled to both vector and hypermultiplets,
with Abelian and non-Abelian gauging, give rise to $\cN=1$ BPS domain
walls connecting different  critical points.

As more specific results, we performed a systematic study of U(1)
gaugings of the toy model with the universal hypermultiplet as well as a
thorough analysis of  a simple model with one vector multiplet and one
hypermultiplet. We studied a family of $\cN=2$ supergravity potentials
with nontrivial vacua that are parametrized by two real numbers. As
another interesting result, this model is found to produce, for $\beta
=-1$ and $\gamma =3/2$, an $\cN=2$ description for the kink solution
of~\cite{FGPW} previously known within the $\cN=8$ theory, and therefore
offers a 2-parameter generalization of this case.   The dual gauge field
theory side of  the models with arbitrary $\beta $ and $\gamma $ is
certainly worthy of investigation.

\

It will be quite natural to apply our apparatus for the search for flows
and critical points in more complicated examples. The first one is a
simple model with no vector but two hypermultiplets \cite{inpreparation},
which has been shown to lead to the RG flow proposed by Girardello,
Petrini, Porrati, and Zaffaroni~\cite{Girardello:1999bd}. The second kind
of example is given by models where non--Abelian gaugings can be
explicitly performed and thus need coupling to more vector multiplets.
Another line of investigation concerns the realization of models dual to
flows from conformal to nonconformal field theories that would need the
presence of both AdS and Minkowski vacua simultaneously in the same model.

A further class of models is the one possessing diverse IR fixed points.
These are aimed at improving the understanding of a possible
supersymmetric realization of the smooth Randall--Sundrum scenario.
Regarding this subject, our work has come to suggest the following
picture. In the presence of hypermultiplets and vector multiplets 5D $\cN
= 2$ gauged  supergravities may have IR AdS${}_5$ fixed points which
eliminate the first reason for the no-go theorem proved in \cite{KL,BC}
for vector multiplets  where only UV critical points exist.

We also found  that the interpolation between 2 IR fixed points (if such
examples are found in the future) for the smooth solution must proceed
through the point where the $W$ superpotential  vanishes. As emphasized
in our discussion, this would not disagree with the monotonicity theorem
for the warp factor $A''\leq 0$. This led us to conclude that a smooth
RSII scenario can take place in the presence of vectors and
hypermultiplets.

On the other hand, the conjectured holographic $c$ theorem is violated
since the $c$ function $c\sim W^{-3}$ blows up at $W=0$. This poses some
problems for the validity of the AdS/CFT correspondence at such points.
However, the general understanding\footnote{This argument was suggested in
discussions with both M. Porrati and L. Susskind.} is that at the
vanishing points of $W$ the physics may not be captured by field theory
but by supergravity and therefore the violation of the holographic $c$
theorem signals that gravity near the wall cannot be replaced by field
theory. This obviously does not happen at $|x^5| \to \infty$ where the
$c$ theorem is expected to be valid and a dual field theory is well
defined.

In this paper we found the general rules for  critical points and zeros
of the superpotential. In more general models, possibly with the use of
other scalar manifolds and different gaugings, one may try to find a
smooth supersymmetric domain wall solution of the RS type.

If the search for a smooth RS scenario remains open, an alternative
strategy would be to introduce some brane sources as in~\cite{susyd5sing}.
This procedure is expected to be quite straightforward as it will require
extending the supersymmetric brane action in a theory with vector
multiplets to include  also the hypermultiplets. A more difficult  step
would be to find the natural mechanism for appearance  of such brane
sources  in string theory with O8-planes and  D8-branes, stabilized
moduli, and additional fluxes along the lines suggested
in~\cite{Bergshoeff:2001pv}.


One could also explore how much of our analysis of 5D can be exported to
4D and 6D, where the geometry described by hypermultiplets is still
quaternionic and where a lot of work has been done for $\cN = 2$
supergravity coupled with vector multiplets only.


Another issue that would be very interesting to discuss is the
11-dimensional origin of the theory at hand. It is indeed known that
five-dimensional supergravity with gauging of the U(1) isometry generated
by $k_1$ can be obtained from $M$-theory compactifications on Calabi--Yau
manifolds in the presence of $G$ fluxes~\cite{Lukas:1998tt,Gflux}.

The same question could be addressed for the new gaugings proposed in
this paper. At first sight their higher dimensional origin seems quite
mysterious, as other isometries are involved, in addition to the shift in
the $\sigma$ scalar field that was discussed in~\cite{Lukas:1998tt,Gflux}.

A related issue is how much of our analysis of domain walls and
supersymmetric vacua may survive in the exact $M$ or string theory rather
than in classical 5D supergravity. The experience with supersymmetric
black hole attractors and quantization of charges suggests the following
possibility. Our domain wall solutions interpolating between
supersymmetric vacua may be hoped to be exact solutions of quantum theory
at most for a restricted  values of gauging parameters.  At the level of
5D classical supergravity these parameters may take  arbitrary continuous
values. Classically, there are no restrictions on the parameters of our
model. Originally, before gauging, they are just parameters of  global
symmetries of ungauged supergravity. These symmetries may well be broken
by quantum effects like instantons. Therefore it would be inconsistent in
the presence of quantum corrections to perform the gauging for continuous
gauging parameters. Only for discrete values of the parameters do we
expect the solutions to be valid when account is taken of quantum
corrections. It is likely  that the clarification of the 11D origin of
the 5D models, taking into account anomalies, fluxes, and quantized
charges of $M$-branes, will shed some light on breaking of continuous
symmetries of gauged supergravities to their discrete subgroups. In such
case the method developed here may provide exact supersymmetric vacua of
$M$ or string theory.

\section*{Acknowledgments}

\noindent We are glad to thank D. Alekseevsky, R. D'Auria, K. Behrndt, E.
Bergshoeff, M. Bianchi, V. Cort{\'e}s, C. Devchand, S. Ferrara, P. Fr{\`e}, G.
Gibbons, C. Johnson, C. Herrmann, M. Kleban, A. Linde, J. Louis, J.
McGreevy, T. Mohaupt, M. Porrati, M. Shmakova, K. Skenderis,  S. Thomas,
L. Susskind,  and S. Vandoren for fruitful discussions. 
The research was supported by the EC under RTN project HPRN-CT-2000-00131,
in which A.C. is associated with Torino University. The work of R.K. was
supported by NSF grant PHY-9870115. A.V.P. thanks the Caltech-USC Center
for hospitality while the paper was finalized.

\newpage
\appendix
\section{Indices}\label{app:indices}
We used in this paper the following indices to describe $n$ vector
multiplets,
 and $r$ hypermultiplets:
\begin{eqnarray}
 \mu  & 0,\ldots ,3,5 & \mbox{local spacetime}     \nonumber\\
 \underline{\mu} & 0,\ldots ,3 & \mbox{4d local spacetime} \nonumber\\
  i  & 1,2 & \mbox{SU(2)-doublets} \nonumber\\
  r & 1,2,3 & \mbox{SU(2)-triplets}\nonumber\\
 I & 0,\ldots ,n & \mbox{vectors} \nonumber\\
 x& 1,\ldots ,n & \mbox{scalars in vector multiplets} \nonumber\\
 A & 1,\ldots ,2r & \mbox{symplectic index for hypermultiplets}\nonumber\\
 X & 1,\ldots ,4r & \mbox{scalars in hypermultiplets} \nonumber\\
\Lambda & 1,\ldots ,n+4r & \mbox{all scalars} \nonumber\\
 \alpha & 1,\ldots ,8& \mbox{SU}(2,1)\mbox{ isometries} \label{indices}
\end{eqnarray}
\section{Reality conditions and SU(2) notations} \label{app:reality}
We first repeat that SU(2) doublet indices $i,j,\ldots $ are raised or
lowered using the NW--SE convention by $\varepsilon _{ij}=\varepsilon
^{ij}$, with $\varepsilon _{12}=1$. The same applies to the Sp(2$r$)
indices $A,B,\ldots $, where a constant antisymmetric matrix $C_{AB}$ is
used, satisfying $C_{AB}C^{CB}=\delta _A{}^C$, with $C^{AB}=(C_{AB})^*$.
By a redefinition, this matrix can be brought into the standard form
$\pmatrix{0&\unity \cr -\unity &0}$. These matrices also enter into
reality conditions. Reality can be replaced by ``charge conjugation''. The
charge conjugation of a scalar [a scalar also in spinor space, an SU(2)
scalar as well as an Sp(2$r$)-scalar] is just its complex conjugate.
Charge conjugation does not change the order of spinors. For a symplectic
Majorana spinor, the charge conjugate is equal to the spinor itself.
However, for a bispinor, one has to introduce a minus sign. Thus, e.g.,
for Majorana spinors $(\bar \lambda\xi )^*=(\bar \lambda\xi )^C=-(\bar
\lambda)^C\xi^C=-\bar \lambda\xi$.

Gamma matrices are ``imaginary'' under this charge conjugation: $\gamma
_a^C=-\gamma _a$. For any object that has SU(2) indices or Sp(2$r$)
indices, the definition of charge conjugation uses the symplectic metric
$(V_i)^C=\varepsilon _{ij}(V_j)^*=(V^i)^*$ and $(V^i)^C=\varepsilon
^{ij}(V^j)^*=-(V_i)^*$, or similarly, $(V_A)^C=C_{AB}(V_B)^*$. All the
quantities that we introduce in the text are real with respect to this
charge conjugation, e.g.,
\begin{equation}
  f^X_{iA}=(f^X_{iA})^C=\varepsilon _{ij}C_{AB}(f^X_{jB})^*\,.
 \label{realvielbeins}
\end{equation}
Symmetric matrices in SU(2) space can be expanded in three components as
\begin{equation}
  R_{(ij)}=\rmi R^r (\sigma _r)_{ij}\, \qquad \mbox{or}\qquad
R^r=\ft12  \rmi R_{(ij)}(\sigma ^r)^{ij} \,.
 \label{RijRr}
\end{equation}
Invariance of $R_{ij}$ under charge conjugation translates into reality of
$R^r$. The usual $\sigma $ matrices are $(\sigma^r)_i{}^j$, and $(\sigma
_r)_{ij}$ is defined from the NW--SE contraction convention: $(\sigma
^r)_{ij}\equiv (\sigma^r)_i{}^k\varepsilon _{kj}$. This leads, e.g., to
$R_{ij}R^{ij}=2R^rR^r$.
\section{Toy model in another parametrization} \label{app:BC}
The manifold $\mathrm{SU}(2,1)/(\mathrm{SU}(2)\times \mathrm{U}(1))$ can
be viewed as an open ball in real 4-dimensional space. Written in complex
coordinates $z_1$ and $z_2$, the domain is $|z_1|^2+|z_2|^2<1$. A useful
parametrization has been introduced in~\cite{Britto-Pacumio:1999sn}, and
used in~\cite{Behrndt:2000ph} to discuss the toy model that we treated in
section~\ref{ss:toymodel}. The variables $z_1$ and $z_2$ are written as
functions of variables $r,\theta ,\varphi,\psi$ as\footnote{The relation
to the variables in section~\ref{ss:scManIsom} is $z_1=(1-S)/(1+S)$ and
$z_2=2C/(1+S)$.}
\begin{equation}
  z_1=r(\cos\ft12\theta) {\rm e}^{\rmi(\psi +\varphi)/2 }\,,\qquad
  z_2=r(\sin\ft12\theta) {\rm e}^{\rmi(\psi -\varphi)/2 }\,.
 \label{z1z2par}
\end{equation}
The manifold is covered by
\begin{equation}
  0\leq r<1\,,\qquad  0\leq \theta<\pi\,, \qquad 0\leq\varphi,\psi<2\pi
  \,.
 \label{region}
\end{equation}
The determinant of the metric is
\begin{equation}
\mathop{\rm det}\nolimits g = \frac{r^6 \sin^2 \theta }{4(1-r^2)^6}\,.
 \label{detg}
\end{equation}
Thus in this parametrization the metric is singular in $r=0$ and for
$\theta =0$. These belong to the manifold, and thus need special care.

In this parametrization, the SU(2) (parameters $\Lambda ^r$) and U(1)
(parameter $\Lambda ^4$) isometries that vanish at the origin take a
simple form on the $z$ variables:
\begin{equation}
  \delta \pmatrix{z_1\cr -z_2} =\pmatrix{-K_{r1}\cr K_{r 2}} \Lambda ^r=
  \ft12\rmi \left[ (\sigma _r)\Lambda ^r +\unity _2 \Lambda
  ^4\right]  \pmatrix{ z_1\cr - z_2}    \,.
 \label{SU2U1}
\end{equation}
We gauge with $K=\alpha_rK_r+\beta K_4$. Apart from the critical point at
the origin, vanishing Killing vectors occur only if there is a zero mode
of the determinant of transformations, i.e., if $|\alpha|=\beta $. We find
two equations:
\begin{eqnarray}
(\alpha_3+\beta)z_1-  (\alpha_1 -\rmi \alpha_2)z_2 &=& 0\,, \nonumber\\
(\alpha_1+\rmi \alpha_2)z_1 +(\alpha_3-\beta) z_2 &=&0\,.
 \label{solz}
\end{eqnarray}
One of the two defines the (real) two-dimensional plane of critical
points, and then the other is automatically satisfied if $|\alpha |=\beta
$. In terms of the angular coordinates, the critical line is at
\begin{equation}
e^{\rmi \varphi }\cot \ft12\theta = \frac{\alpha_1+\rmi
\alpha_2}{\alpha_3+\beta}\,,\qquad
 e^{2\rmi \varphi }=\frac{\alpha_1-\rmi
\alpha_2}{\alpha_ 1+\rmi \alpha_2}\,,\qquad
  \cot^ 2  \ft12\theta = \frac{(\alpha_1)^2+(\alpha_2)^2}{(\alpha_3+\beta)^2}\,.
 \label{critlineang}
\end{equation}
Although there is clearly no difference in the choice of the direction in
SU(2) space, the choice of angular coordinates makes the gauging in the
direction $\alpha_3$ difficult. For example, the Killing vectors in the
angular coordinates are
\begin{eqnarray}
 K_1 & = & (0,- \sin \varphi ,- \cos \varphi  \cot\theta ,
   \cos\varphi \sin^{-1} \theta )\,, \nonumber\\
 K_2 & =  & (0,- \cos\varphi ,  \sin\varphi\cot\theta   ,-  \sin\varphi  \sin^{-1}
 \theta)\,,\nonumber\\
 K_3 &=& (0,0,-1,0)\,,\nonumber\\
 K_4&=& (0,0,0,-1)\,.
   \label{Kparangles}
\end{eqnarray}
This gives the impression that $\alpha_3 K_3 +\beta K_4$ never vanishes,
not even at $r=0$! However, in this case, the two combinations of $z$
mentioned above are $z_1=0$ and $z_2=0$. The latter is the line $\theta
=0$ where the parametrization degenerates! Note that these singularities
are coordinate singularities. There is nothing generically different for
gauging in the direction ``3,'' as this direction is equivalent to the
others in the symmetric space. The different features that are mentioned
in~\cite{Behrndt:2000ph} are artifacts of the parametrization, which is
singular at $r=0$ and at $\theta =0$. It is precisely at $\theta = 0$
that these authors obtain different results from ours.

\begin{figure}
\begin{center}
\leavevmode \epsfxsize=5cm
 \epsfbox{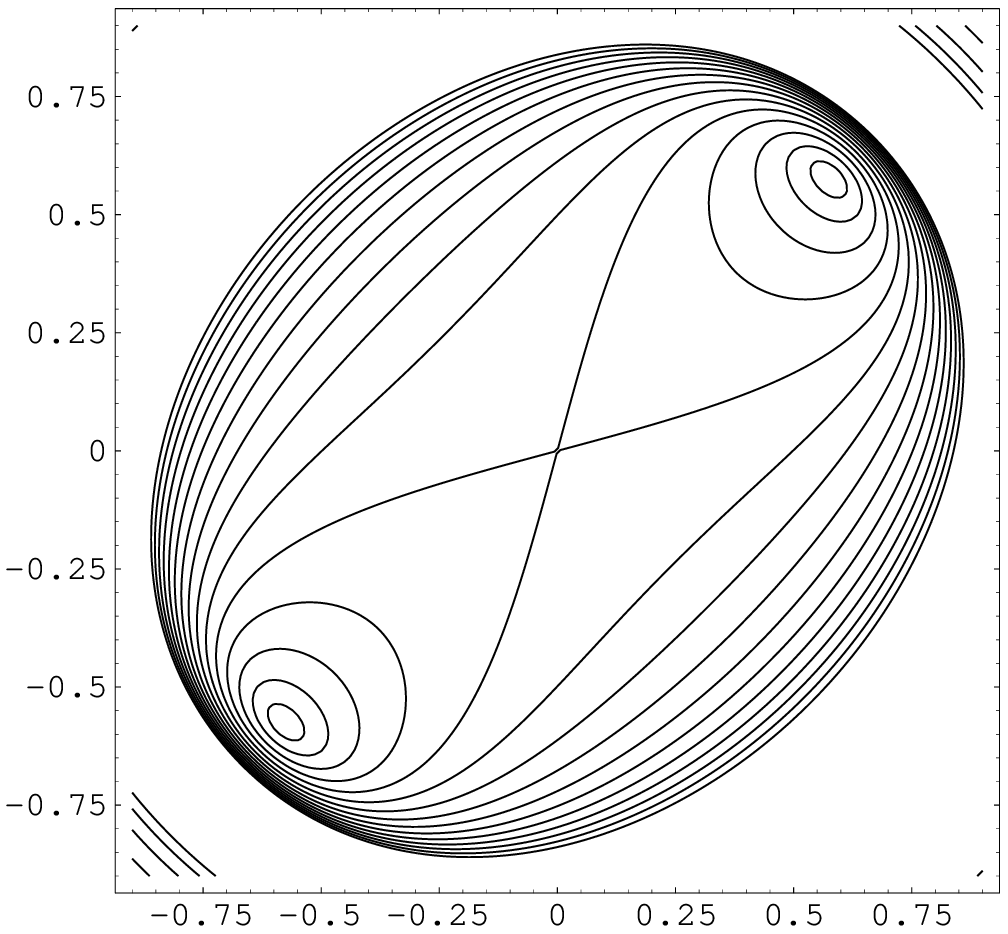}\hspace{3cm}\epsfxsize=5cm
 \epsfbox{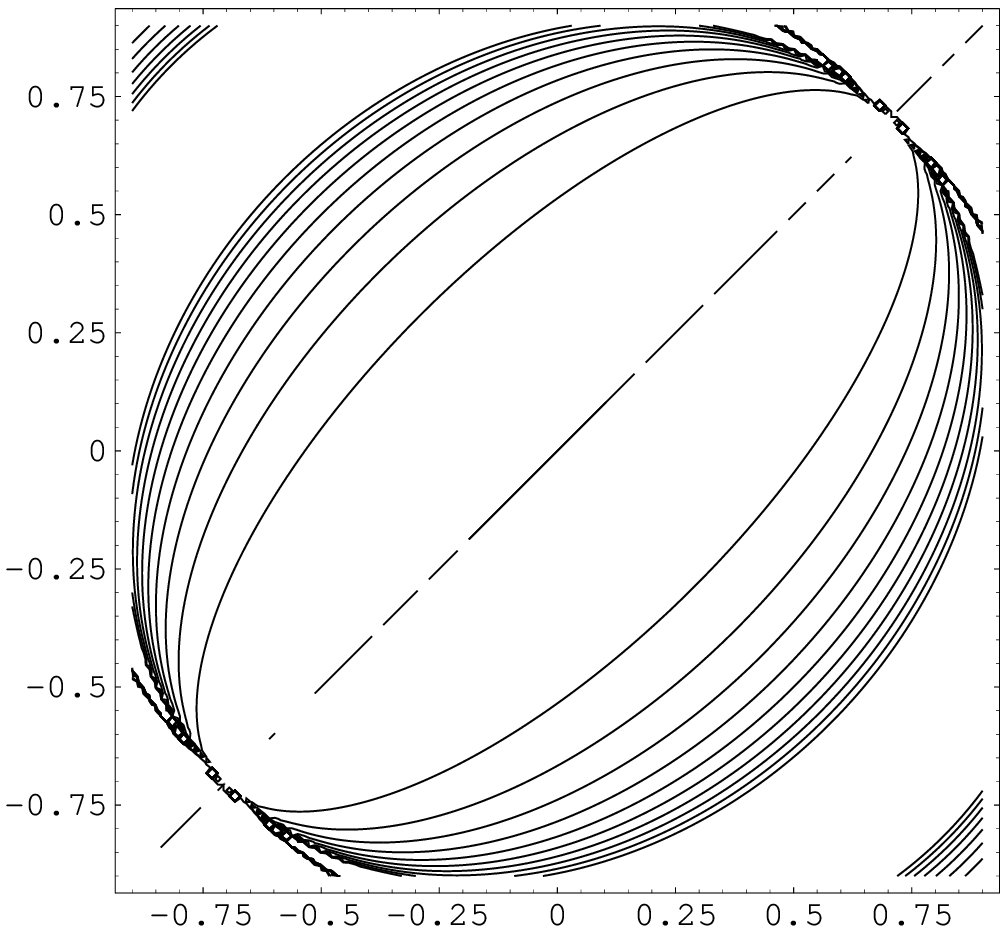}
\caption{\it Contours of constant $W$ in the plane $(\Re z_1,\Re z_2)$
for $\alpha_1=\sqrt{3/2}$ and $\beta=2\sqrt{3/2}$ (left) and for
$\alpha_1=\beta=\sqrt{3/2}$ (right). \label{fig:contplotW}}
\end{center}
\end{figure}

To avoid the singularities, and for showing the main features, we will
further concentrate on gauging in direction ``1'' for the SU(2) and the
U(1) direction; thus
\begin{equation}
  \alpha_2=\alpha_3=0\,, \qquad \alpha_1\beta>0\,.
 \label{dir14}
\end{equation}
With the latter choice, the critical line is at $\varphi =0$, $\theta
=\pi /2$, or $z_1=z_2$. The zeros that we mentioned in
section~\ref{ss:toymodel} occur now for
\begin{equation}
   \varphi =0\,,\qquad \theta =\ft12 \pi \,,\qquad
  r^2=\frac{2\alpha_1}{\alpha_1+\beta}\,.
 \label{zeroW}
\end{equation}
This point is only part of the domain if $\beta>\alpha_1$. The Killing
vector is nonzero at such points. In this case, the nonzero component is
\begin{equation}
  K_\psi =\beta - \alpha_1\,.
 \label{Kpsi}
\end{equation}
The total prepotential can be written as
\begin{equation}
  W^2= \frac{(\beta r^2)^2+2\alpha_1\beta(r^2-2)\zeta +(\alpha_1)^2(4-4r^2+\zeta
  ^2)}{6(1-r^2)^2}\,,
 \label{W2zeta}
\end{equation}
where we use
\begin{equation}
  \zeta \equiv r^2 \sin\theta \cos\varphi =z_1\bar z_2+\bar z_1z_2\,.
 \label{defzeta}
\end{equation}
As $W$ depends just on two real parameters, we can plot it in the plane
for real $z_1$ and $z_2$ to see the whole picture. This leads to the
contour plot in figure~\ref{fig:contplotW}
for a typical case $\beta>\alpha_1$ (left figure). Observe that it is
similar to figure~\ref{fig:conpl34}, which represented the gauging in
direction ``3'' in the other representation. The crucial line is the
diagonal and along this line the potential is again the one of
figure~\ref{fig:potbc}. We also clearly see that the line $\theta =0$
(horizontal line in the graph) does not have any special properties. The
critical points that were found there in~\cite{Behrndt:2000ph} came out
of the analysis only due to the singular nature of the parametrization.

For the case $\beta=\alpha_1$ we have
\begin{equation}
  W=\sqrt{\frac{2}{3}}\left| \beta\frac{1-\ft12|z_1+z_2|^2}{1-|z_1|^2-|z_2|^2}\right|\,.
 \label{W2a1isa4}
\end{equation}
The potential is then constant on the line $z_1=z_2$, and we have the
right plot in figure~\ref{fig:contplotW}. The culmination points of the
lines are at $r=1$, i.e., they do not belong to the manifold.

This establishes the equivalence of the two parametrizations. In
particular, only one critical point or connected set of critical points
is possible.
\newpage

\providecommand{\href}[2]{#2}\begingroup\raggedright\endgroup

\end{document}